\documentclass[aps,prd,superscriptaddress,showpacs,amsmath,amssymb,amsfonts,nofootinbib,showpacs,reprint]{revtex4-1}

\pdfoutput=1

\usepackage{graphicx}
\usepackage{color}
\usepackage{xcolor}
\usepackage{slashed}
\usepackage{tabularx}
\usepackage{hyperref}
\usepackage{ifthen}
\usepackage{multirow}
\usepackage{makecell}
\usepackage{float}
\usepackage{microtype}
\usepackage[english]{babel}

\def\simge{\mathrel{
     \rlap{\raise 0.511ex \hbox{$>$}}{\lower 0.511ex \hbox{$\sim$}}}}
\def\simle{\mathrel{
     \rlap{\raise 0.511ex \hbox{$<$}}{\lower 0.511ex \hbox{$\sim$}}}}
\def\be{\begin{equation}}
\def\ee{\end{equation}}
\def\bea{\begin{eqnarray}}
\def\eea{\end{eqnarray}}

\renewcommand{\Re}{{\rm Re}\,}
\renewcommand{\Im}{{\rm Im}\,}

\newcommand{\beq}{\begin{eqnarray}}
\newcommand{\eeq}{\end{eqnarray}}

\newcolumntype{L}[1]{>{\raggedright\arraybackslash}p{#1}} 
\newcolumntype{C}[1]{>{\centering\arraybackslash}p{#1}} 
\newcolumntype{R}[1]{>{\raggedleft\arraybackslash}p{#1}} 

\renewcommand{\bar}[1]{\mkern 1.5mu\overline{\mkern-1.5mu#1\mkern-1.5mu}\mkern 1.5mu}

\graphicspath{{.}{figures/}{mma/}{figures}}

\definecolor{blue}{rgb}{0,0,1}

\definecolor{green}{rgb}{0,1,0}

\definecolor{red}{rgb}{1,0,0}

\begin{document}

\title{QCD Anderson transition with overlap valence quarks on a twisted-mass sea}

\newcommand{\JLU}{Institut f\"ur Theoretische Physik, Justus-Liebig-Universit\"at, Heinrich-Buff-Ring 16, 35392 Giessen, Germany}
\newcommand{\HFHF}{Helmholtz Forschungsakademie Hessen f\"ur FAIR (HFHF), GSI Helmholtzzentrum f\"ur Schwerionenforschung, Campus Gießen}
\newcommand{\FAIR}{Facility for Antiproton and Ion Research in Europe GmbH (FAIR GmbH), 64291 Darmstadt, Germany}

\author{Robin Kehr}\affiliation{\JLU}
\author{Dominik Smith}\affiliation{\JLU}\affiliation{\FAIR}
\author{Lorenz von Smekal}\affiliation{\JLU}\affiliation{\HFHF}

\begin{abstract}
In this work we probe the QCD Anderson transition by studying spectral distributions of the massless overlap operator on gauge configurations created by the \emph{twisted mass at finite temperature collaboration} (tmfT) with 2+1+1 flavors of dynamical quarks and the Iwasaki gauge action. We assess finite-size and discretization effects by considering two different lattice spacings and several physical volumes, and mimic
the approach to the continuum limit through stereographic projection. Fitting the inflection  points of the participation ratios of the overlap Dirac eigenmodes, we obtain estimates of the temperature dependence of the mobility edge, below which quark modes are localized. We observe that it  is well-described by a quadratic polynomial and systematically vanishes at temperatures below the pseudocritical one of the chiral transition. 
In fact, our best estimates within errors overlap with that of the chiral phase transition temperature of QCD in the chiral limit. 
\end{abstract}

\maketitle

\section{Introduction}\label{sec:intro}

Since Anderson's seminal work on the localization properties of non-interacting electrons in systems with random disorder \cite{Anderson:1958vr}, Anderson localization has been studied abundantly in condensed matter systems \cite{Evers:2007zsx}. In the Anderson tight-binding model the Hamiltonian contains a random on-site potential which produces
an exponential localization of electron eigenmodes by destructive quantum interference \cite{FROHLICH19849,2008arXiv0807.2531M}.
In more than two dimensions, there is a critical strength of the disorder potential at which all eigenmodes get localized, and a metal-insulator transition characterized by a vanishing conductivity occurs.
For disorder strengths below this critical one an energy threshold emerges, known as the \textit{mobility edge}, which separates 
the spectrum of the Hamilton operator into localized and delocalized regimes.

An intriguing idea, originally inspired by the instanton-liquid model \cite{GarciaGarcia:2005dp,GarciaGarcia:2005vj} and later 
supported by numerical evidence from lattice QCD \cite{GarciaGarcia:2006gr}, is that a similar transition occurs in finite-temperature QCD (and other QCD-like theories), whereby the eigenmodes of the Dirac operator 
play the role of the electron modes in Anderson's model. 
Within lattice gauge theory this has since been extensively 
studied, beginning with the $\mathrm{SU}(2)$ gauge theory
~\cite{Bruckmann:2008xr,Kovacs:2009zj,Kovacs:2010wx,Nishigaki:2013uya}, 
as well as in three-color lattice QCD, quenched \cite{Kovacs:2017uiz,Kovacs:2018mxg,Giordano:2019pvc}
and unquenched 
\cite{Gavai:2008xe,Kovacs:2012zq,Nishigaki:2013uya,Giordano:2013taa,Giordano:2014pfa,Ujfalusi:2015nha}. 
In each case, the spectrum of the Dirac 
operator in the high-temperature phase exhibits a localized and a 
delocalized regime, separated by a mobility edge $\lambda_{\mathrm c}$.
In quenched QCD, the onset of Anderson localization coincides with
the deconfining phase transition \cite{Kovacs:2017uiz,Kovacs:2018mxg,Giordano:2019pvc}, as it does in 
a lattice model for QCD with three degenerate light flavors of unimproved staggered fermions on coarse lattices, where there is an artificial first-order phase transition in the Polyakov loop \emph{and} the chiral condensate \cite{Pittler:2014qea,Giordano:2016nuu}.
This suggests a close relationship between the three phenomena, 
and that temperature might control the amount of disorder 
in some analogy to the disorder parameter in the (unitary) Anderson model, cf.~Refs.~\cite{Giordano:2018iei,Giordano:2021qav,Bruckmann:2011cc}. 

For a rather recent review on localization in gauge theories, see \cite{Giordano:2021qav}. Numerical evidence for the localization of low-lying Dirac modes preferentially in regions where disorder emerges in the Polyakov loop was given in \cite{Bruckmann:2011cc,Cossu:2016scb,Holicki:2018sms,Baranka:2021san,Baranka:2022dib}.  The connection between localized near-zero modes and the possible existence of massless single-particle Goldstone excitations in the chiral limit is discussed in detail in \cite{Giordano:2022ghy}. The overall picture emerging is as follows: An accumulation of possibly localized or critical near-zero modes, as observed right above the pseudocritical temperature in lattice QCD with physical and smaller than physical light-quark masses \cite{Dick:2015twa,Ding:2020xlj,Kaczmarek:2021ser}, might delocalize, as the critical temperature in the chiral limit is approached from above, to produce the chiral condensate and Goldstone pions with vanishing mobility edge right at the chiral phase transition.  
The disappearance of the mobility edge  and Anderson localization at the chiral phase transition in QCD can therefore in principle provide no less than the connection between chiral symmetry restoration and deconfinement. 

In this work, we study the spectrum and eigenmodes of the massless overlap operator,
which exactly fulfills the Ginsparg-Wilson relation,
on the background of $\mathrm{SU}(3)$ gauge field configurations generated by 
the \emph{twisted mass at finite temperature collaboration} (tmfT) \cite{Burger:2013hia,Burger:2015xda,Burger:2017xkz}
with $N_\mathrm{f}=2+1+1$ flavors of twisted-mass Wilson quarks at maximal twist and using the Iwasaki 
gauge action. Such a mixed-action setup was first proposed in Refs.~\cite{Cichy:2010ta,Cichy:2012vg} 
where its feasibility was demonstrated. 
The first use of such a setup to investigate Anderson localization was reported in Ref.~\cite{Holicki:2018sms}. 
Our present study improves upon the previous one by considering larger physical volumes and smaller lattice spacings.
The focus is a more precise determination of the temperature dependence of the mobility edge, using the inverse participation ratio of Dirac eigenmodes. 

Some earlier works have reported an approximately linear dependence $\lambda_\mathrm{c}(T) \approx c\,(T-T_0)$ \cite{Pittler:2014qea,Holicki:2018sms}, although some evidence of curvature in this relation had already been reported in \cite{Kovacs:2012zq}.
Compared to the preliminary previous study within the same setup \cite{Holicki:2018sms}, here we observe that the quadratic contribution can be extracted more reliably, when the effects of the lattice spacing and volume, and of unphysically large bare quark-masses are controlled. Using a stereographic projection of eigenvalues from the Ginsparg-Wilson circle to the imaginary axis to mimic the continuum limit, we then obtain evidence that the mobility edge vanishes very close to current estimates of the critical temperature of the chiral phase transition in the $m_\mathrm{q} \to 0$ limit. This contrasts with previous studies, which have placed the Anderson transition closer to the pseudocritical temperature, i.e.~in the proximity of the $m_\mathrm{q} \neq 0$ crossover. 

This work is structured as follows: We start by describing our lattice setup in detail in Sec. \ref{sec:setup}, including a comprehensive list of the lattice configuration which were analyzed. In Sec. \ref{sec:observables} we then describe our target observables, namely the distribution of eigenvalues and the inverse participation ratio, and our methods for obtaining them. 
We present results in Sec. 
\ref{sec:results}. This includes 
the raw eigenvalue spectrum to illustrate the formation of a Banks-Casher gap, as well as our study of the temperature dependence of the mobility edge. The section is split into two parts: We first discuss the dependence of all observables on the polar angle of the Ginsparg-Wilson circle, and then in terms of the purely imaginary eigenvalues after stereographic projection. We summarize and conclude in Sec. \ref{sec:conclusion}.

\section{Lattice setup}\label{sec:setup}

In this work, we use lattice configurations created by the \emph{twisted mass at finite temperature collaboration} under guidance of the late Michael M\"uller-Preussker \cite{Burger:2013hia,Burger:2015xda,Burger:2017xkz}. 
The ensembles were generated with $N_\mathrm{f}=2+1+1$ flavors of dynamical twisted-mass Wilson fermions at maximal twist \cite{Frezzotti:2003xj}, comprising two 
degenerate light plus physical strange and charm quarks, and using the Iwasaki gauge action.
These configurations are a subset of the ones presented in Ref.~\cite{Burger:2018fvb}
and are listed comprehensively in Table 
\ref{tab:qcd_anderson_lattice_setup_overview}. We adopt the nomenclature of Ref. \cite{Burger:2018fvb}, as well as the values of the pion mass in physical units, lattice spacing and pseudocritical temperatures listed therein. The scale setting and tuning of the twist parameter is based on earlier zero-temperature simulations reported in Ref.~\cite{Alexandrou:2014sha} and in the work by the \emph{European twisted mass collaboration} (ETMC) \cite{Baron:2008xa, Baron:2009zq, Baron:2010bv, Baron:2011sf}.

We employ a fixed scale approach. 
Our spectral computations were performed on three sets of ensembles (denoted by A370, D370 and D210) with two different lattice spacings $a$, for several temperatures fixed by $N_\mathrm{t}$. 
While strange and charm quarks have physical masses, the pion 
mass is still unphysically large in the ensembles considered here, with $m_\pi = 364(15)\,\mathrm{MeV}$, $m_\pi = 369(15)\,\mathrm{MeV}$ and $m_\pi = 213(9)\,\mathrm{MeV}$, respectively. The chiral crossover occurs at $T_\mathrm{pc} \approx 185\,\mathrm{MeV}$ for larger pion masses and at $T_\mathrm{pc}=158(5)\,\mathrm{MeV}$ for the smaller one.
The physical sizes $L = a N_\mathrm s$ of the spatial volumes are such that $m_\pi L \approx 4.14$ (A370), $3.86$ (D370), and $3.34$ (D210). 
Therefore, relative to the Compton wavelength of the pion the physical volume in the D210 ensemble is at the lower end and residual finite volume effects are to be expected, although with $N_\mathrm s = 48$  the largest spatial lattice size (of $L \simeq 3.1$~fm) was used in this case.  

\begin{table}
          \centering
          \begin{tabular}{|c|c|c|c|c|c|c|}
          \hline
          Set of ensembles   &   $N_\mathrm{s}$   &   $N_\mathrm{t}$   &   $T$ / $\mathrm{MeV}$ &   $T/T_\mathrm{pc}$ &   \# $\mathrm{conf.}$   &   $\frac{\mathrm{modes}}{\mathrm{conf.}}$                                                  \\
          \hline
          \multirow{9}{*}{\makecell{\textbf{A370} \\ 
          $a = 0.0936(13)$\,fm    \\ $m_\pi=364(15)\,\mathrm{MeV}$ \\ $T_\mathrm{pc}=185(8)\,\mathrm{MeV}$}}
                                                
          &   \multirow{9}{*}{24} &   4   &   527(7)  &   2.85(13)    &   200     &   200         \\
          &                       &   5   &   422(6)  &   2.28(10)    &   200     &   160         \\
          &                       &   6   &   351(5)  &   1.90(9)     &   200     &   135         \\
          &                       &   7   &   301(4)  &   1.63(7)     &   150     &   115         \\
          &                       &   8   &   264(4)  &   1.42(6)     &   200     &   100         \\
          &                       &   9   &   234(3)  &   1.27(6)     &   200     &   90          \\
          &                       &   10  &   211(3)  &   1.14(5)     &   250     &   80          \\
          &                       &   11  &   192(3)  &   1.04(5)     &   200     &   75          \\
          &                       &   12  &   176(2)  &   0.95(4)     &   200     &   70          \\
          \hline
          \multirow{6}{*}{\makecell{\textbf{D370} \\ 
          $a = 0.0646(7)$\,fm    \\ $m_\pi=369(15)\,\mathrm{MeV}$ \\ $T_\mathrm{pc}=185(4)\,\mathrm{MeV}$}}
                                                
          &   \multirow{4}{*}{32} &   3   &   1018(11)    &   5.50(13)    &   120     &   400     \\
          &                       &   6   &   509(6)      &   2.75(7)     &   120     &   200     \\
          &                       &   14  &   218(2)      &   1.18(3)     &   160     &   85      \\
          &                       &   16  &   191(2)      &   1.03(2)     &   160     &   75      \\
          \cline{2-7}
          &   40                  &   18  &   170(2)      & 
          0.92(2)     &   40      &   150     \\
          \cline{2-7}
          &   48                  &   20  &   153(2)      &   0.83(2)     &   3       &   200     \\
          \hline
          \multirow{5}{*}{\makecell{\textbf{D210} \\ 
          $a = 0.0646(7)$\,fm  \\ $m_\pi=213(9)\,\mathrm{MeV}$ \\ $T_\mathrm{pc}=158(5)\,\mathrm{MeV}$}}
          &   \multirow{5}{*}{48} &   4   &   764(8)  &   4.83(16)    &   10  &   1000            \\
          &                       &   6   &   509(6)  &   3.22(11)    &   10  &   700             \\
          &                       &   8   &   382(4)  &   2.42(8)     &   10  &   500             \\
          &                       &   10  &   305(3)  &   1.93(6)     &   10  &   400             \\
          &                       &   12  &   255(3)  &   1.61(5)     &   10  &   350             \\
          \hline
          \end{tabular}
          \caption{List of tmfT ensembles used in this work, with numbers of configurations on which overlap eigenmodes were computed and numbers of eigenmodes per configuration. Parameters and nomenclature adopted from Refs.~\cite{Burger:2018fvb} and \cite{Alexandrou:2014sha}. Due to excessive computational costs, the eigenmodes for the D370 ensemble with $N_\mathrm{s}=48$ and $N_\mathrm{t}=20$ suffer from low statistics.}
          \label{tab:qcd_anderson_lattice_setup_overview}
\end{table}

For our measurements we employ the overlap Dirac operator \cite{Neuberger:1997fp, Neuberger:1998wv, Neuberger:1998my, Neuberger:2003nu}. In particular, we study the spectrum and eigenmodes of the massless overlap operator 
\begin{equation}
D = \frac{1}{\tilde a}\left( 1 + \mathrm{sgn}\,K \right)\,, \label{eq:Dov}
\end{equation}
which exactly fulfills the Ginsparg-Wilson relation 
$\left\{ \gamma_5, D \right\} = \tilde a D \gamma_5 D$ \cite{Ginsparg:1981bj,Luscher:1998pqa}, 
where $\tilde a = a/\rho $ is proportional to the lattice spacing $a$. 
$D$ obeys an Atiyah-Singer index theorem, implements the $U_\mathrm{A}(1)$ anomaly on the lattice and is automatically $\mathcal O(a)$ improved \cite{Niedermayer:1998bi,Capitani:1999uz}.
The sign function $\mathrm{sgn}\,K = \frac{K}{\sqrt{K^\dag K}}$ is computed using a rational (Zolotarev) approximation, while for the kernel $K$ we used the Wilson 
operator with negative mass parameter $-\rho$ 
where $\rho \in (0,2)$ is a scaling factor which can be tuned to optimize the locality properties of the overlap operator. 
We set it to $\rho=1.4$ in this work as determined in Ref.~\cite{Cichy:2012vg} to be an optimal choice, if no additional smearing of the gauge fields is employed.\footnote{The rescaling $\tilde a = a/\rho $ in \eqref{eq:Dov} then restores the correct tree-level form of the massless overlap operator in the continuum limit.}
 
\section{Observables}\label{sec:observables}

\subsection{Eigenvalue spectrum}
      \label{sec:spectrum}
      
The spectral density $\rho(\lambda)$ of a chiral Dirac operator encodes spontaneous chiral symmetry breaking via
the \textit{Banks-Casher relation} \cite{Banks:1979yr,DeGrand:2006nv,Fukaya:2010na} which connects the near-zero spectral density of the Dirac operator with the chiral condensate $\langle\overline\psi \psi \rangle $ used as the order parameter for chiral symmetry breaking and defined after suitable additive renormalization, e.g.~via strange condensate or connected susceptibility subtraction in lattice QCD \cite{Unger:2010wcq}, 
\begin{equation}
                        \langle \overline\psi \psi \rangle =  -\pi \lim_{\lambda\to0} \lim_{m\to0} \lim_{V\to\infty} \rho(\lambda)\,.
                        \label{eq:banks_casher_relation}
\end{equation}
At low temperatures, $\langle \overline\psi \psi \rangle \propto \lim_{\lambda\to0}\,\rho(\lambda) >0$, while chiral symmetry restoration in turn implies $\lim_{\lambda\to0}\,\rho(\lambda) =0$.
This implies that a gap opens in the spectral density of the Dirac operator above $T_\mathrm{pc}$.
At high temperatures, and at sufficient distance from the edge of the gap, the spectral density of eigenstates shows an 
approximate square root behavior \cite{Damgaard:2000cx}. 
More recently, it has been argued that another thermal phase transition exists in QCD above the chiral crossover transition \cite{Alexandru:2019gdm,Alexandru:2021pap}. This transition was estimated to occur at $200 \text{MeV} < T_\mathrm{IR} < 250 \text{MeV} $ and involves a proliferation of deep infrared near-zero modes with $0 \lesssim \lambda\ll T $. In the pure gauge theory these near-zero modes were found to encode the decoupling of scale-invariant long-range correlations in the infrared \cite{Alexandru:2019gdm}.
Although there are  near zero modes also in this so-called IR phase,
the corresponding scale-invariant infrared correlations cannot contribute to the order parameter for chiral symmetry breaking, i.e.~the subtracted chiral condensate in the Banks-Casher relation. 

The eigenmodes and eigenvalues of the massless overlap operator are the primary target of this study.
To diagonalize the overlap operator, we use the \textit{Krylov-Schur method} \cite{krylov-schur}, 
as provided by the SLEPc library \cite{slepc}. In previous work \cite{Holicki:2018sms}, the \textit{implicitly restarted Arnoldi method} (IRAM) provided by the ARPACK library was employed instead \cite{iram,ARPACKUserGuide}. However, it was found that the Krylov-Schur method yields a significantly better performance, which was tested for one configuration per $N_\mathrm{t}$ of the D210 data set. Additionally, we noticed that one has to be careful using the ARPACK++ library \cite{arpack++}, which provides a C++ interface to ARPACK with 32-bit integers, since integer overflows can occur for large scale diagonalizations, as done for this work.\footnote{The SLEPc library, on the other hand, can be installed with a 64-bit integer option and contains such an interface as well.}  

\begin{figure}
   \centering
                              
\includegraphics[width=0.385\textwidth]{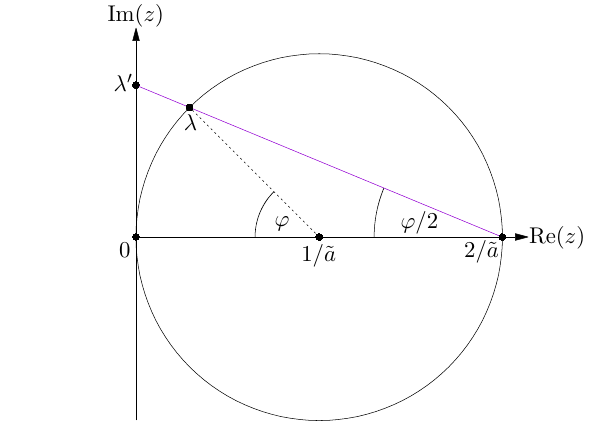}
\includegraphics[width=0.433\textwidth]{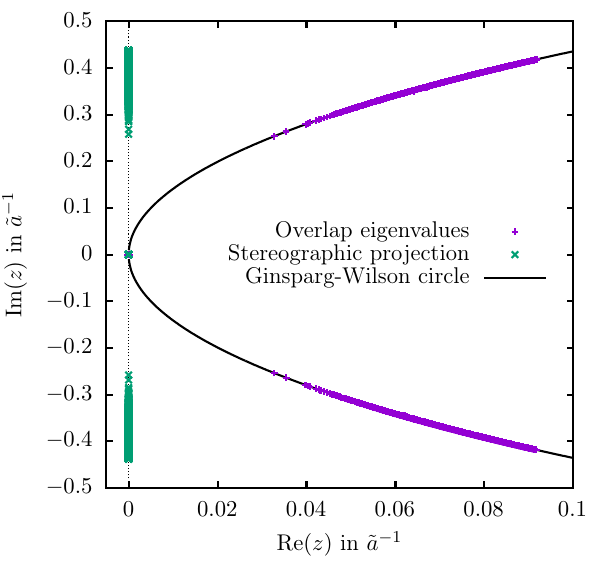}
 		\caption{\textit{Top}: Stereographic projection of the eigenvalues at finite $\tilde a= a/\rho $ to the imaginary axis. 
	\textit{Bottom}: Projection of eigenvalues at $T=509.10(5.52)\,\mathrm{MeV}$ for the D210 data set.}
    \label{fig:qcd_anderson_stereographic_projection_example}
\end{figure}

In the continuum, the Dirac eigenvalues are 
purely imaginary, but on the lattice the overlap eigenvalues $\lambda $ are distributed on the Ginsparg-Wilson circle with radius $1/\tilde a$, given by
  \begin{equation}
  \Im \lambda = \pm\,\sqrt{ \frac{2}{\tilde a}\,\Re \lambda - \left( \Re \lambda \right)^2 }\,. \label{eq:Ginsparg-Wilson}
                        \end{equation}
In this work, we first present a discussion of the raw complex eigenvalues in terms of their polar angle, 
 \begin{equation}
 \varphi = \arctan \left ( \frac{\Im \lambda}{\tilde a^{-1}-\Re \lambda} \right) \,, \label{eq:varphi}
 \end{equation}
on the Ginsparg-Wilson circle centered at $1/\tilde a$. In order to relate the overlap eigenvalues $\lambda $ to the purely imaginary ones of the (massless) Dirac operator in the continuum, as in Ref.~\cite{Bietenholz:2003mi}, we then stereographically project the Ginsparg-Wilson circle onto the imaginary axis, see Fig.~\ref{fig:qcd_anderson_stereographic_projection_example}, and repeat the analysis with the projected ones for all data sets as a rough mock of a continuum extrapolation, below.

\subsection{Inverse participation ratio}
\label{sec:ipr}

To probe and quantify localization, we use the relative space-time volume $r(\lambda)$ that is occupied by an eigenmode. This relative volume or participation ratio $r(\lambda) $ is computed using the inverse participation ratio (IPR) $P_2(\lambda)$. 
The $q$-th order (generalized) IPR of an eigenmode $v_\lambda$, see, e.g., Refs.~\cite{Evers:2007zsx,GarciaGarcia:2006gr,Gavai:2008xe,Ujfalusi:2015nha}, is defined as the eigenstate moment $P_q$,
\begin{equation}
  P_q(\lambda)  = \sum_{i \in \Lambda} ( v_\lambda(i)^\dagger  v_\lambda(i))^q\,, 
\end{equation}
where the sum runs over all sites $i$ of the $d+1$ dimensional lattice $\Lambda$. The volume dependence
of these moments is such that for extended (metallic) modes they behave as $P_q \sim L^{-d(q-1)}$, while for exponentially localized modes $P_q\sim 1 $. They are separated by critical modes which have non-trivial fractal dimensions $0<D_q <d $ and $P_q \sim L^{-D_q(q-1)}$. 
		
Because the inverse of the second moment $P_2$ is by definition proportional to the (spacetime) volume that the eigenmode occupies, one can define the relative eigenmode volume
\begin{equation}
			r(\lambda) = \frac{P_2^{-1}(\lambda)}{|\Lambda|} \in [1/|\Lambda|,1]\,, \label{eq:EigenModeVol}
\end{equation}
where $r(\lambda) \sim L^{-d} $ and $r(\lambda) \sim 1$ correspond to localized and delocalized modes, respectively. For critical modes, $r(\lambda) \sim L^{-\eta}  $, where $\eta = d - D_2$ determines the spatial correlations of the probability density $|v_\lambda(i)|^2$ \cite{Evers:2007zsx}.
For the unitary Anderson model in $d=3$ dimensions, it was numerically determined as $\eta \simeq 1.827 $ \cite{2015PhRvB..91r4206U}, for example.

In practice the eigenvalues are discrete, and $r(\lambda)$ scatters even for nearby values of $\lambda$, so we average $r(\lambda)$ over small bins of size $\Delta \lambda$ to approximate a smooth function. The exact procedure is as follows: For each lattice configuration of a given parameter set, we compute a number of the lowest
eigenvalues as quoted in Table \ref{tab:qcd_anderson_lattice_setup_overview} and measure $r$ on the corresponding eigenmodes. For each bin, we then obtain the mean value and
standard error of $r$-measurements from all configurations, for which $\lambda$ falls into it. In principle, this might introduce correlations both within a single bin and between different bins, as there might exist subtle dependencies between eigenmodes of each individual configuration, even if different configurations are statistically independent. To rule out the first case, we have confirmed for all results shown in this work, that a bin-wise Jacknife error analysis reproduces the standard error. The second case is a bit more involved, and is discussed in more detail in Appendix~\ref{sec:appendix1}. The main conclusion of the analysis presented there, however, is that the residual correlations of this second kind play no significant role for the main conclusions of this qualitative study of the mobility edge separating the localized from the extended Dirac eigenmodes.

For a proper finite-size-scaling analysis, in principle, one should then analyze $L^\eta \, r(\lambda) $ in different spatial volumes, 
which approaches zero(infinity) for localized(extended) modes, and only stays finite for critical ones, in the infinite volume limit.
One could then determine the mobility edge $\lambda_\mathrm{c}$ from an extrapolation of the pairwise intersection points of successive volumes. In lack of the necessary data for a suitable number of volumes, here we simply approximate $\lambda_\mathrm{c}$ by the spectral position of the inflection point of the bin-averaged relative participation ratio $\bar{r(\lambda)}$ which tends to $\lambda_\mathrm{c}$ in the infinite volume limit.

\section{Results}
\label{sec:results}
\subsection{Angle dependence}
\label{sec:results_angle_dependence}

\begin{figure}
    \centering
    \includegraphics[width=0.47\textwidth]{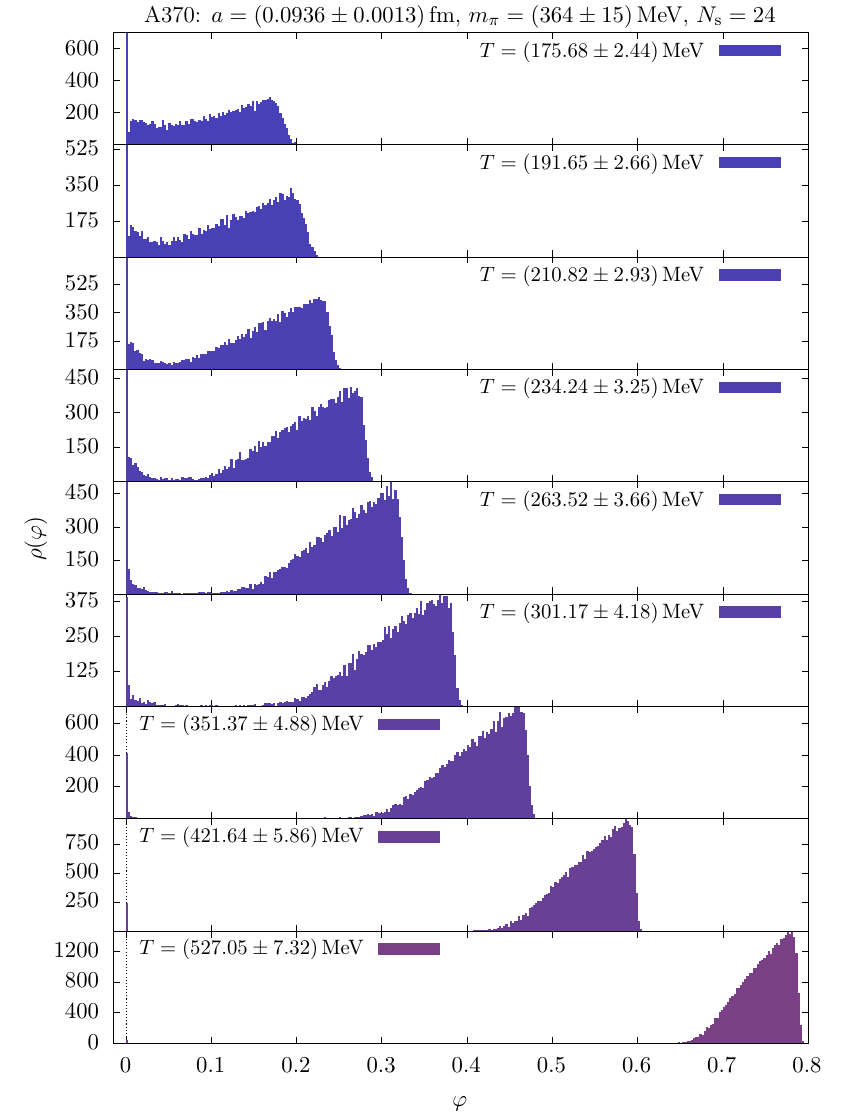}
    \caption{Histogram of the polar angle distribution $\rho(\varphi)$ with bin size $\Delta\varphi = 2.5\cdot10^{-3}$ of the overlap eigenvalues for the A370 configurations ($a=0.0936(13)\,\mathrm{fm}$; $m_\pi = 364(15)\,\mathrm{MeV}$). \label{fig:qcd_anderson_dist_A370_angle}}
\end{figure}

\begin{figure}
    \centering
    \includegraphics[width=0.47\textwidth]{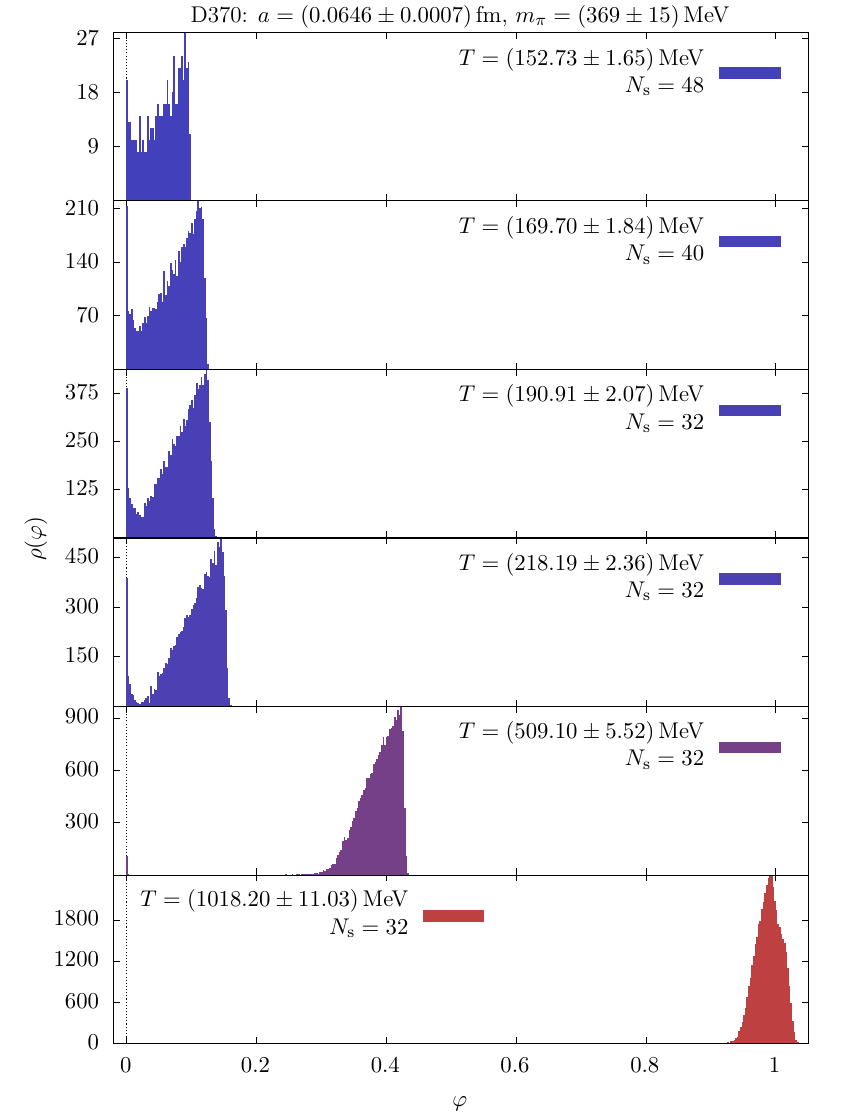}
    \vspace{-2mm}
    \caption{Same as in Fig.~\ref*{fig:qcd_anderson_dist_A370_angle} but for the D370 configurations ($a=0.0646(7)\,\mathrm{fm}$; $m_\pi = 369(15)\,\mathrm{MeV}$).}     \label{fig:qcd_anderson_dist_D370_angle}
    \vspace{4mm}
    \includegraphics[width=0.47\textwidth]{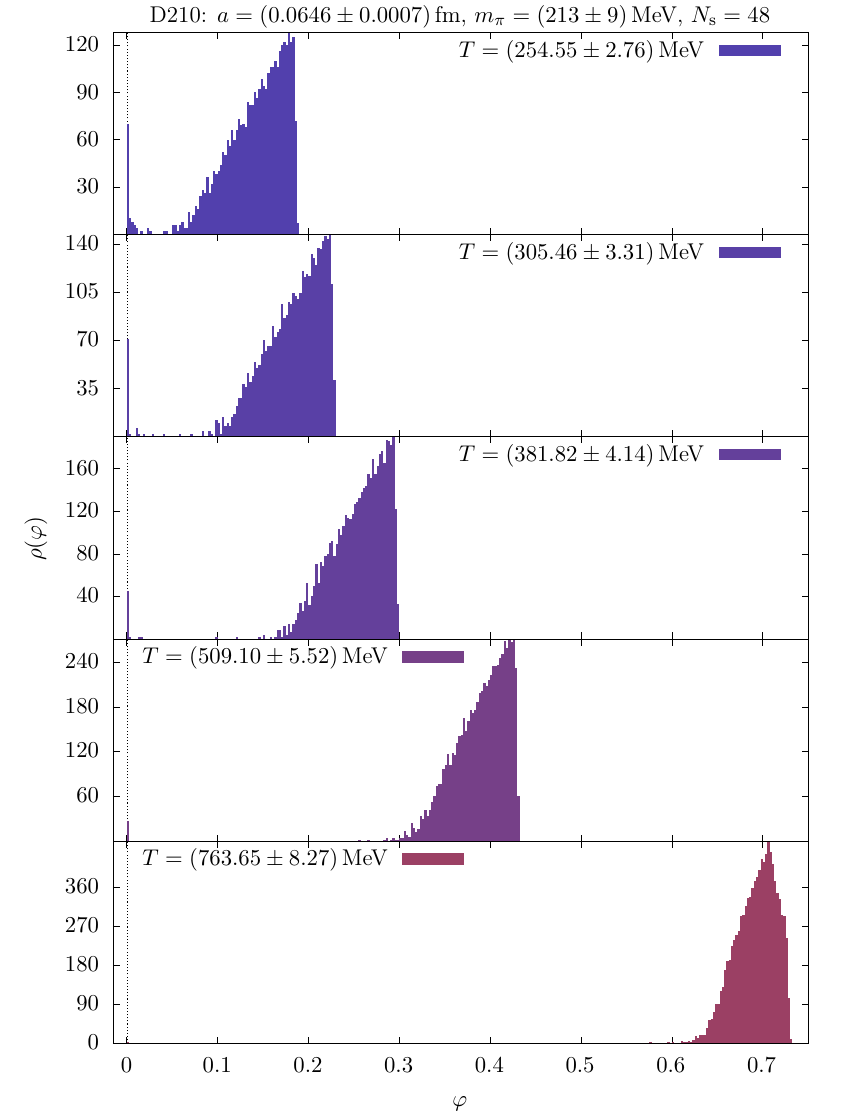}
    \caption{Same as in Fig.~\ref*{fig:qcd_anderson_dist_A370_angle} but for the D210 configurations ($a=0.0646(7)\,\mathrm{fm}$; $m_\pi = 213(9)\,\mathrm{MeV}$).}
\label{fig:qcd_anderson_dist_D210_angle}
\end{figure}

We start by discussing the Anderson transition in terms of the polar angle $\varphi$, from Eq.~\eqref{eq:varphi}, of the complex eigenvalues of the overlap operator which are distributed on the Ginsparg-Wilson circle. Since the eigenvalues come in complex conjugate pairs with equal relative volume, it is sufficient to consider the absolute value of the eigenvalues. For better readability, we redefine $\varphi \equiv \bar{|\varphi|}$ and $r(\varphi)\equiv \bar{r(\varphi)}$ in the following discussion. In Figs.~\ref{fig:qcd_anderson_dist_A370_angle}, 
\ref{fig:qcd_anderson_dist_D370_angle} and 
\ref{fig:qcd_anderson_dist_D210_angle} we show 
histograms with bin size $\Delta\varphi = 2.5\cdot10^{-3}$, which estimate the angle-dependent eigenvalue probability density $\rho(\varphi)$
for the A370, D370 and D210 data set for different temperatures.

The chiral transition occurs at $T_\mathrm{pc} \approx 185\,\mathrm{MeV}$ (for the A370 and D370 data set) and at  $T_\mathrm{pc} \approx 158\,\mathrm{MeV}$ (D210) respectively (see Table \ref{tab:qcd_anderson_lattice_setup_overview}); most of the presented temperatures are thus in the chirally symmetric phase. We clearly observe the Banks-Casher gap, which grows as the temperature increases. 
Due to finite volume and discretization effects, there are still eigenmodes within the Banks-Casher gap at small temperatures (close to $T_\mathrm{pc}$) where it has not fully opened yet. Whether the accumulation of near zero modes observed at temperatures around $T \approx 250$~MeV and below are lattice artifacts or an indication of the infrared modes of an emerging IR phase as predicted in Refs.~\cite{Alexandru:2021pap,Alexandru:2019gdm}, remains to be studied more carefully on finer lattices and in larger volumes, in the future. In addition, each ensemble also contains a number of exact zero modes with definite chirality.

\begin{figure}
    \centering
     \includegraphics[width=0.47\textwidth]{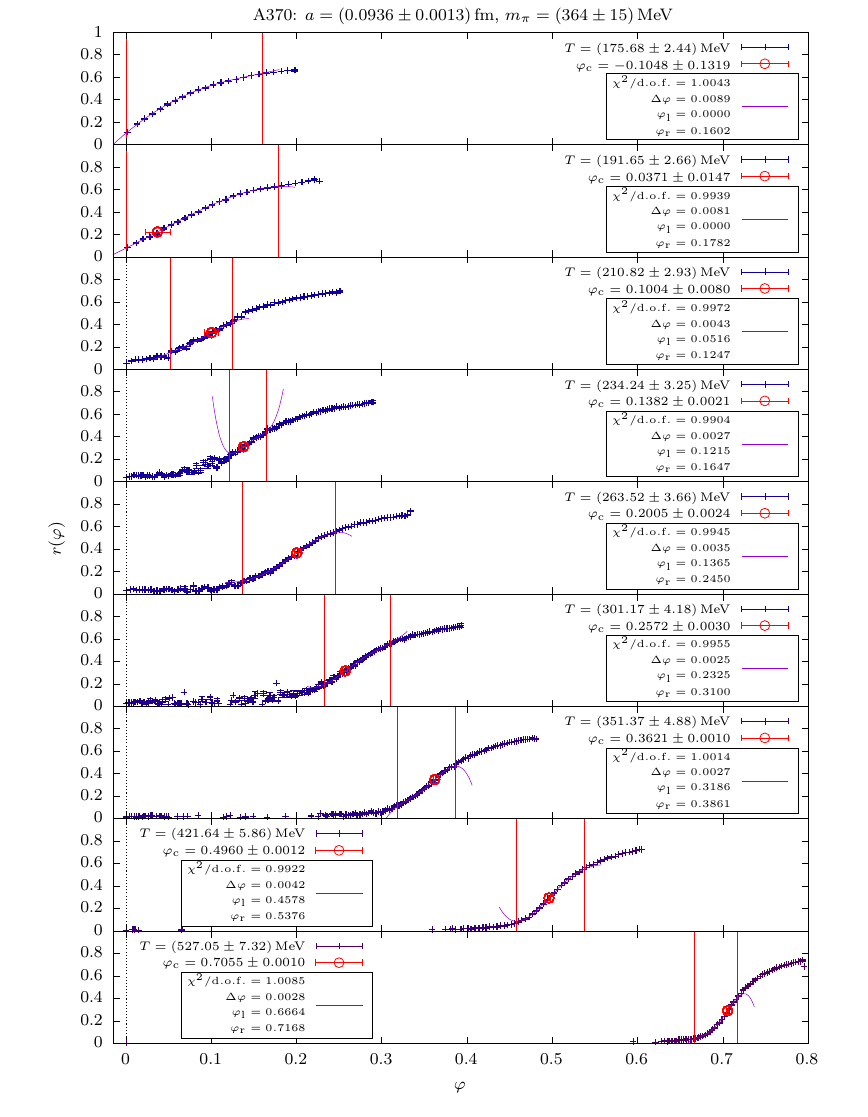}
     \vspace{-2mm}
    \caption{Bin averages of the relative eigenmode volume $r(\varphi)$ defined in Eq.~(\ref*{eq:EigenModeVol}), with $\Delta\varphi = 2.5\cdot10^{-3}$, over the polar angle of the overlap eigenvalues, with fits~(\ref*{eq:inflectionfit}), fit windows and extracted inflection  points marked in red, for the A370 configurations.          
\label{fig:qcd_anderson_ipr_A370_angle}}
%
    \vspace{4mm}
     \includegraphics[width=0.47\textwidth]{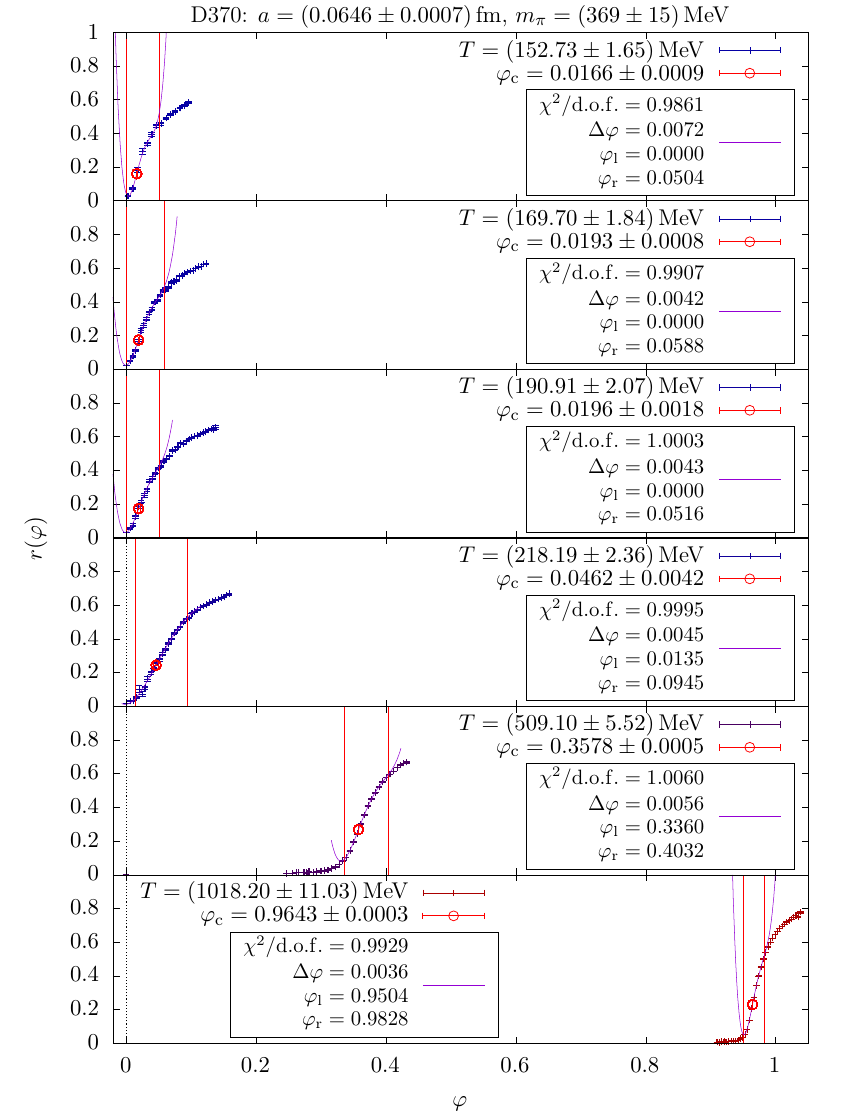}
    \caption{Same as in Fig.~\ref*{fig:qcd_anderson_ipr_A370_angle} but
    for the D370 configurations.
\label{fig:qcd_anderson_ipr_D370_angle}}
\end{figure}
                
\begin{figure}
    \centering
     \includegraphics[width=0.49\textwidth]{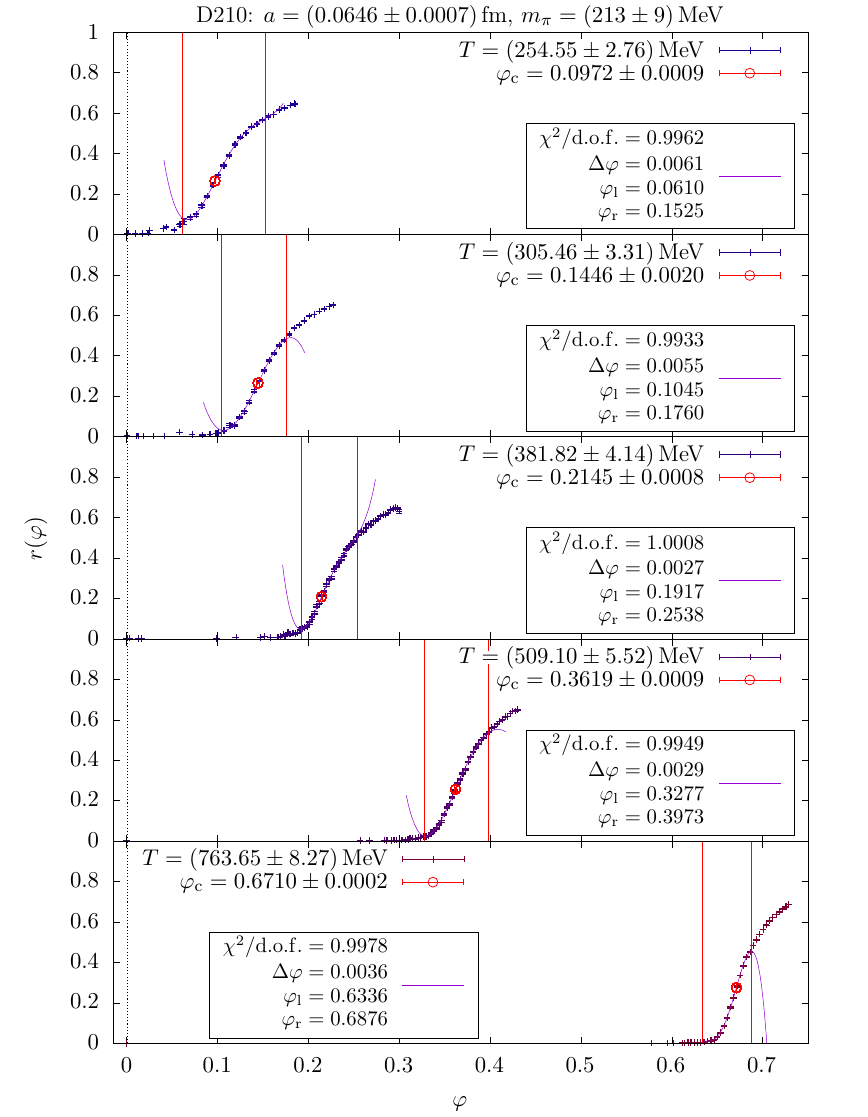}
     \caption{Same as in Fig.~\ref*{fig:qcd_anderson_ipr_A370_angle} but for the D210 configurations.
\label{fig:qcd_anderson_ipr_D210_angle}}
\end{figure}

In Figs.~\ref{fig:qcd_anderson_ipr_A370_angle},  \ref{fig:qcd_anderson_ipr_D370_angle} and \ref{fig:qcd_anderson_ipr_D210_angle} the relative eigenmode volume $r(\varphi)$ is shown for each data set.  
We estimate the mobility edge by determining the inflection points $\varphi_\mathrm{c}$, hence
		\begin{equation}
			\frac{\partial^2 r(\varphi)}{\partial \varphi^2} \bigg|_{\varphi=\varphi_\mathrm{c}} = 0\,,
		\end{equation}
using fits of the form 
		\begin{equation}
 r(\varphi)= r_\mathrm{c}+b(\varphi-\varphi_\mathrm{c})+c(\varphi-\varphi_\mathrm{c})^3+d(\varphi-\varphi_\mathrm{c})^4\,.
 \label{eq:inflectionfit}
		\end{equation} 
To maximize the goodness of fit of our model, aside from the fit constants $r_\mathrm{c} $, $\varphi_\mathrm{c} $, $b$, $c$, $d$,
we vary the lower (left) and upper (right) bounds of the fit-window $\varphi_\mathrm{l}$ and $\varphi_\mathrm{r}$, and the bin size $\Delta\varphi$ over which $r(\varphi)$ is averaged. In doing so, we obtain  fits where $\chi^2 / \mathrm{d.o.f.}$ is unity to at least one digit precision for most of the data sets. 
The optimized bin size and  boundaries as well as $\chi^2 / \mathrm{d.o.f.}$
are given in boxes as part of the legend in each figure. The corresponding fit windows are furthermore indicated by vertical red lines, and the resulting inflection points $\varphi_\mathrm{c}$ as red circles (with errorbars). With the single exception of the lowest temperature in the A370 ensemble (where a value consistent with zero
is obtained for  $\varphi_\mathrm{c}$), the inflection point always falls well inside the fit window.

In fact, the pertinent fits are of such quality that they are hard to distinguish from the data inside the fit windows in our figures.
For better visibility, we have therefore plotted these fits slightly beyond the respective windows used to obtain them. The so obtained values of the inflection points $\varphi_\mathrm{c}$ as our final estimates of the mobility edge are listed in Table~\ref{tab:qcd_anderson_lambdac_from_inflection_point_angular}.

\begin{figure}
    \centering                        \includegraphics[width=0.5\textwidth]{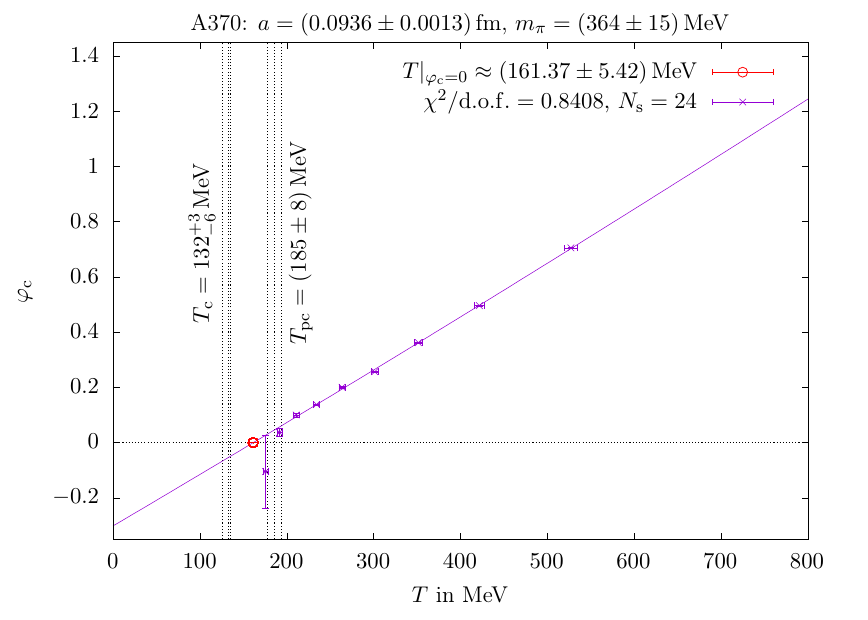}\\
    \vspace{2mm}
    \includegraphics[width=0.5\textwidth]{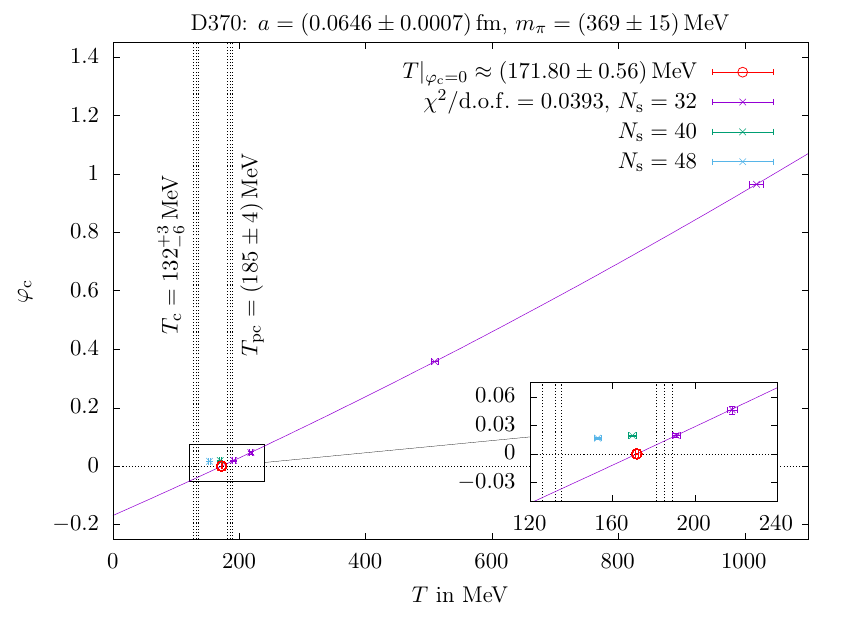}\\
    \vspace{2mm}
    \includegraphics[width=0.5\textwidth]{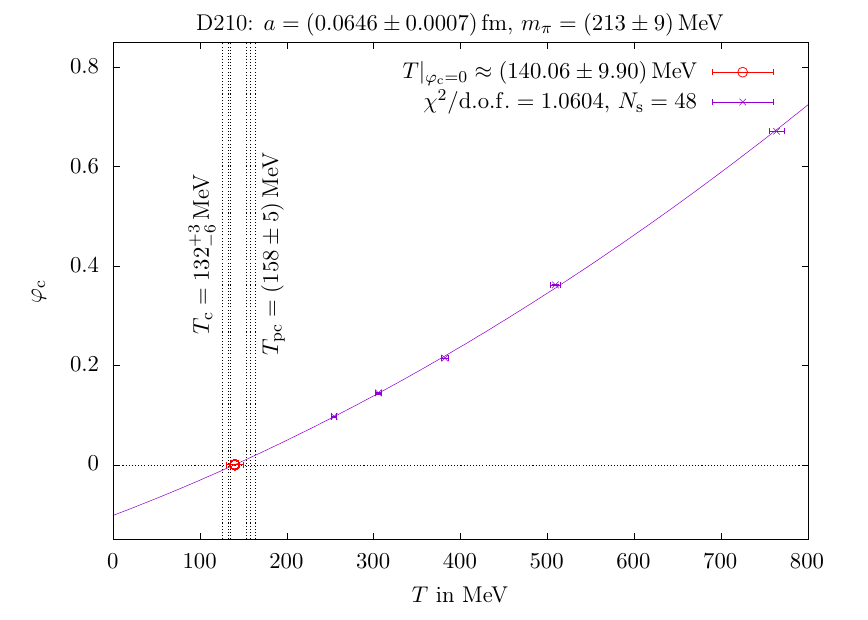}\\
    \caption{Temperature dependence of mobility edge $\varphi_\mathrm{c}(T)$ for the A370 (top), D370 (middle) and D210 (bottom) configurations, as extracted from the bin-averaged angular relative eigenmode volumes in Figs.~\ref*{fig:qcd_anderson_ipr_A370_angle},  \ref*{fig:qcd_anderson_ipr_D370_angle} and \ref*{fig:qcd_anderson_ipr_D210_angle}, with fits (\ref*{eq:MobilityEdgeFit}) and $T_0$ extrapolations (red) where $\varphi_\mathrm{c}(T_0) = 0$.
\label{fig:qcd_anderson_mobility_edge_over_temperature_angle}}
\end{figure}

We then use these values to estimate the Anderson transition temperature $T_0$, where the mobility edge vanishes, applying fits of the form
 \begin{equation}
		    \varphi_\mathrm{c}(T)=b(T-T_0) +c(T-T_0)^2\,,
      \label{eq:MobilityEdgeFit}
		\end{equation}
to all data sets. We thereby use the uncertainties in the lattice spacings to obtain error bars in $T$, which we then apply to all data points in the fitting procedure. The resulting curves are shown in Fig.~\ref{fig:qcd_anderson_mobility_edge_over_temperature_angle} together with the raw data $\varphi_\mathrm{c}(T)$ obtained from the inflection points for the A370 (top), D370 (middle) and D210 (bottom) ensembles, respectively. Our estimates of $T_0$, where $\varphi_\mathrm{c}=0$, are indicated by red circles together with the fit error bars (which hardly exceed the symbol size)  in these plots.

The temperature dependence of the mobility edge from the coarse and small lattice with heavier pion mass used for the A370 data set is approximately linear as observed previously \cite{Holicki:2018sms}. We have evaluated two more sets of configurations with $N_\mathrm{t} = 13$ and $14$, both with $N_\mathrm{s}$ = 32 corresponding to larger physical volumes. While these indicate that finite volume effects play no major role for the coarse A370 data set, at least qualitatively, they were nevertheless excluded from the fit due to the different volume, however.
As for the $N_\mathrm{t} = 12$ data (shown in the top of Figure \ref{fig:qcd_anderson_ipr_A370_angle}), independent of the volume, no mobility edge is observed at these lower temperatures either. The respective data is therefore not relevant here and neither listed in Table~\ref{tab:qcd_anderson_lattice_setup_overview} nor shown in the figures.
For the D370 data set we find values of $\chi^2 / \mathrm{d.o.f.} \ll 1$, indicating over-fitting, which is likely due to the comparatively small number of data points: The two sets of configurations with $N_\mathrm{t} = 18$ and $20$, which are listed as part of the D370 data set in Table \ref{tab:qcd_anderson_lattice_setup_overview}, have larger $N_\mathrm{s}$ values (hence larger physical volumes) and were excluded from the fit, thus leaving only four data points for a three parameter fit. Displayed separately in the figures, the values of $\varphi_\mathrm{c}$ for these two data sets from the larger spatial volumes do indicate, however, that $T_0$ will tend to values smaller than the extrapolated one from the fit (indicated by the red circle) when the  volume is increased. 
Therefore, finite volume effects appear to become relevant again closer towards the continuum limit in a fixed volume. 

While this discussion suggests that a linear model would suffice to describe the rather sparse data (from the same volume) in this case as well, the quadratic term is certainly necessary for the D210 data set, on the other hand. In comparison, this data set is the most important one, however, because it has the largest physical volume, smallest lattice spacing and closest to physical pion mass, anyway. With these most realistic lattice parameters, the curvature is then clearly visible in the bottom plot of Figure~\ref{fig:qcd_anderson_mobility_edge_over_temperature_angle}. 

Moreover, it is important to note here, that in all cases we obtain $T_0$ estimates which fall just outside (in particular below) the error bands of the pseudocritical temperatures $T_\mathrm{pc}$ as determined for the chiral crossover transition in Ref.~\cite{Burger:2018fvb} with these ensembles, and indicated together with their respective error bands by the vertical dashed lines labeled $T_\mathrm{pc}$ in these figures. 
For comparison, we have also included the estimate of the critical temperature $T_\mathrm{c}$ of the chiral phase transition (for $m_\mathrm{q} \to 0$) taken from Ref.~\cite{HotQCD:2019xnw} in the same way. This estimate is confirmed by the more recent one from \cite{{Kotov:2021rah}}. 
Incidentally, in the D210 case, we find that our $T_0$ error bar slightly overlaps with the $T_\mathrm{c}$ error band for the chiral phase transition temperature.

		\begin{table}
			\centering
			\begin{tabular}{|c|c|c|c|c|c|}
					\hline
						Set of ensembles &	$N_\mathrm{s}$   &   $N_\mathrm{t}$	&	$T$ / MeV	&  $T/T_\mathrm{pc}$	&  $\varphi_\mathrm{c}$ \\
					\hline
					\multirow{9}{*}{\makecell{\textbf{A370} \\ $T_0 = 161(5)\,\mathrm{MeV}$}}
					& \multirow{9}{*}{24}
						& 4		&	527(7)	&  2.85(13)  &  0.7055(10)		\\
					&	& 5		&	422(6)	&  2.28(10)  &	0.4960(12)		\\
					&	& 6		&	351(5)	&  1.90(9)   &	0.3621(10)	    \\
					&	& 7		&	301(4)	&  1.63(7)   &	0.2572(30)		\\
					&	& 8		&	264(4)	&  1.42(6)   &	0.2005(24)		\\
					&	& 9		&	234(3)	&  1.27(6)   &	0.1382(21)		\\
					&	& 10	&	211(3)	&  1.14(5)   &	0.1004(80)	    \\
					&	& 11	&	192(3)	&  1.04(5)   &	0.0371(147)		\\
					&	& 12	&	176(2)	&  0.95(4)   &	-0.1048(0.1319)		\\
					\hline
					\multirow{6}{*}{\makecell{\textbf{D370} \\$T_0 = 172(1)\,\mathrm{MeV}$}}
					&\multirow{4}{*}{32}
						  &	 3	&	1018(11) &	5.50(13)  &  0.9643(3)	    \\
					&	  &	 6	&	509(6)	 &	2.75(7)   &  0.3578(5)	    \\
					&	  &	14	&	218(2)	 &	1.18(3)   &  0.0462(42)	    \\
					&	  &	16	&	191(2)	 &	1.03(2)   &  0.0196(18)	    \\
					\cline{2-6}
					&  40 & 18	&	170(2)	 &	0.92(2)   &  0.0193(8)   \\
                    \cline{2-6}
					&  48 &	20	&	153(2)	 &	0.83(2)   &  0.0166(9)	    \\
					\hline
					\multirow{5}{*}{\makecell{\textbf{D210}\\$T_0 = 140(10)\,\mathrm{MeV}$}}
					&\multirow{5}{*}{48}
					   &  4  & 764(8)  &  4.83(16)  &  0.6710(2)            \\
                    &  &  6  & 509(6)  &  3.22(11)  &  0.3619(9)            \\
                    &  &  8  & 382(4)  &  2.42(8)   &  0.2145(8)            \\
                    &  & 10  & 305(3)  &  1.93(6)   &  0.1446(20)           \\
                    &  & 12  & 255(3)  &  1.61(5)   &  0.0972(9)            \\
					\hline
			\end{tabular}
            \vspace{2mm}
   			\caption{Angular mobility edges $\varphi_\mathrm{c}$ determined from the inflection points in our fits~(\ref*{eq:inflectionfit}) to the bin-averaged relative eigenmode volumes $r(\varphi)$ in Figs.~\ref*{fig:qcd_anderson_ipr_A370_angle},  \ref*{fig:qcd_anderson_ipr_D370_angle} and \ref*{fig:qcd_anderson_ipr_D210_angle}, together with the extrapolated $T_0$ values from Fig.~\ref*{fig:qcd_anderson_mobility_edge_over_temperature_angle}.}
\label{tab:qcd_anderson_lambdac_from_inflection_point_angular}
		\end{table}
		
\subsection{Stereographic projection}

As a means of estimating, at least qualitatively, the effect of  continuum extrapolations, we apply the stereographic projection sketched in Fig.~\ref{fig:qcd_anderson_stereographic_projection_example}, from the Ginsparg-Wilson circle onto the imaginary axis, to the eigenvalues for all data sets. It corresponds to the M\"obius transformation 
$f(z) = z/ (1-\tilde a z/2) $, which maps the Ginsparg-Wilson circle onto the imaginary axis and is uniquely determined by $f(z) = z + \mathcal O(\tilde a z^2) $ and $f(2/\tilde a) = \infty $ \cite{Bietenholz:2003mi}. This amounts to defining the purely imaginary quantity 
\begin{equation}
    \lambda^\prime:= f(\lambda) = \frac{\mathrm{i}\, \Im \lambda}{1 - \frac{\tilde a}{2} \Re \lambda} 
    =  \mathrm{i}\, \text{\scalebox{.9}{$\frac{2}{\tilde a}$}} \tan\text{\scalebox{.9}{$\frac{\varphi}{2}$}} ~,
\end{equation}
and repeating the investigation of the last section in terms of this new variable. Again, we redefine $\lambda:=\bar{|\lambda^\prime|}$ and $r(\lambda):=\bar{r(\lambda)}$ for better readability in the following.

\begin{figure}
\centering
\includegraphics[width=0.47\textwidth]{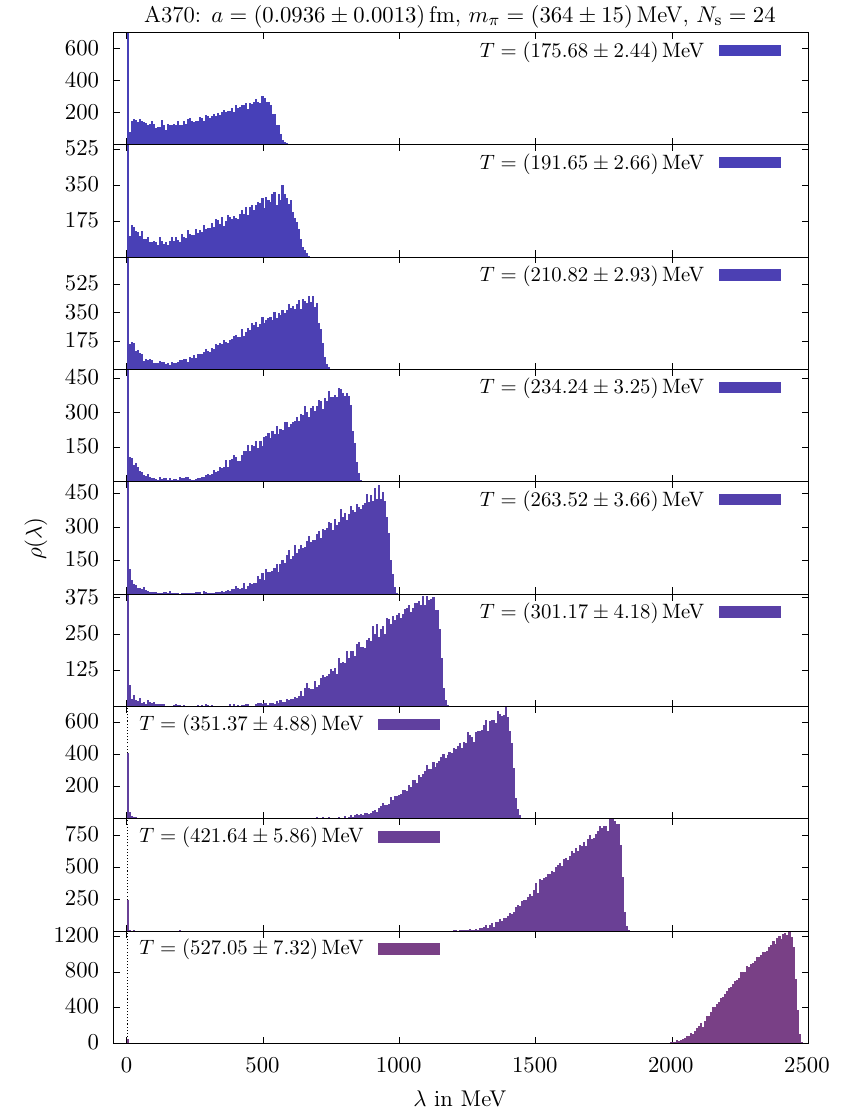}
\vspace{-2mm}
\caption{
Histogram of the distribution $\rho(\lambda)$ of the stereographically projected overlap eigenvalues $\lambda$ for the A370 configurations ($a=0.0936(13)\,\mathrm{fm}$; $m_\pi = 364(15)\,\mathrm{MeV}$).    \label{fig:qcd_anderson_dist_A370}}
\vspace{2mm}
\includegraphics[width=0.47\textwidth]{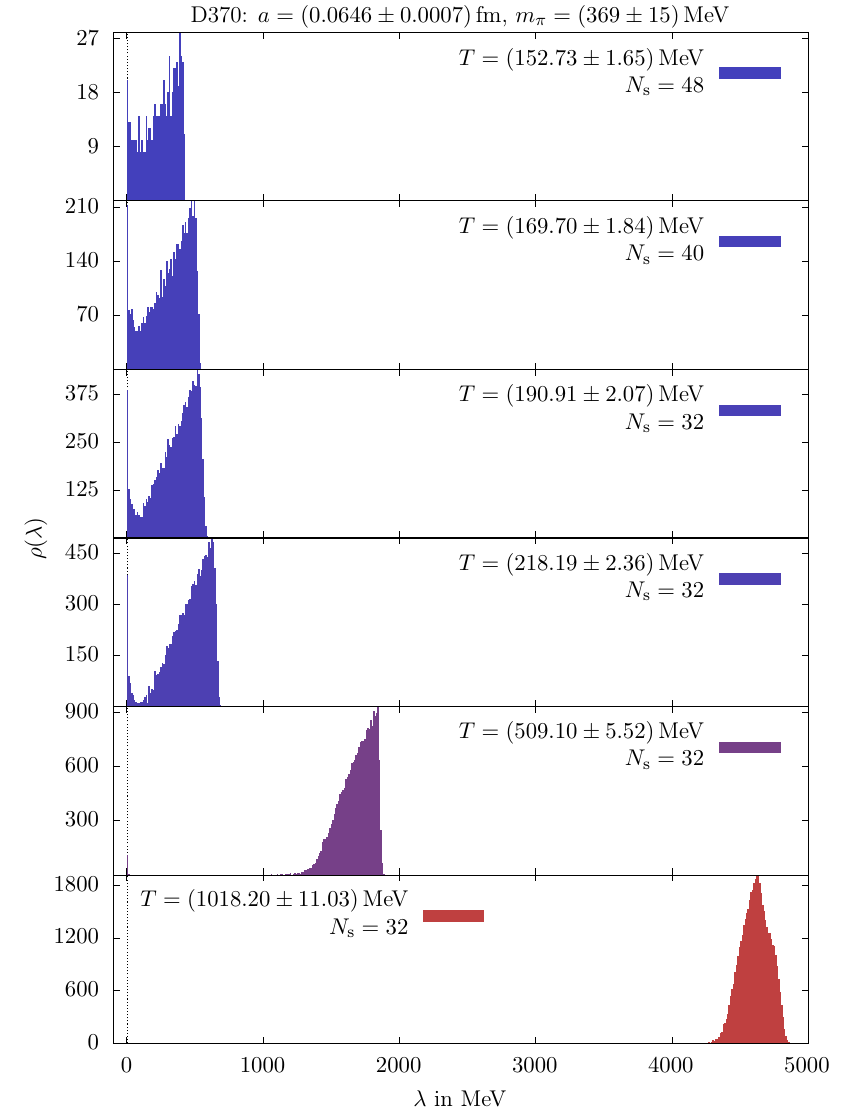}
\vspace{-2mm}
\caption{Same as in Fig.~\ref*{fig:qcd_anderson_dist_A370}, here for the D370 configurations ($a=0.0646(7)\,\mathrm{fm}$; $m_\pi = 369(15)\,\mathrm{MeV}$).  \label{fig:qcd_anderson_dist_D370}}
\end{figure}

\begin{figure}[t]
    \centering
    \includegraphics[width=0.47\textwidth]{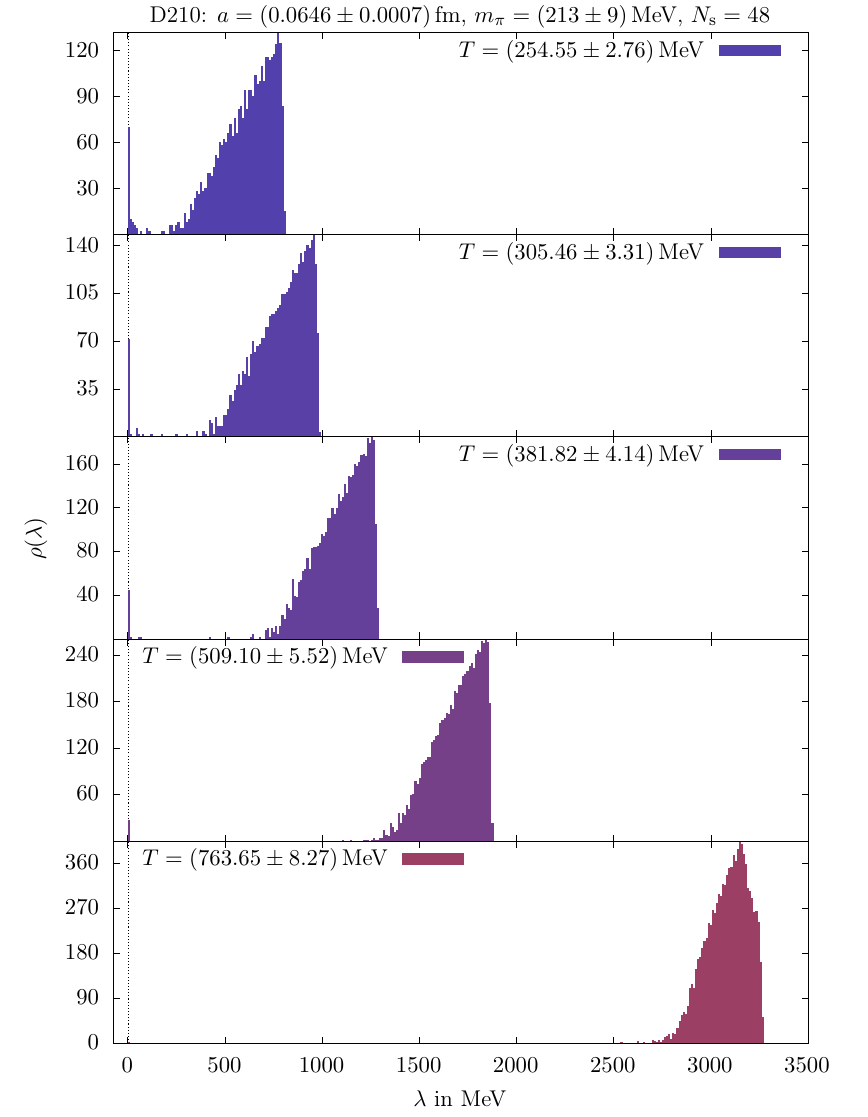}
\caption{Same as in Fig.~\ref*{fig:qcd_anderson_dist_A370}, here for the D210 configurations ($a=0.0646(7)\,\mathrm{fm}$; $m_\pi = 213(9)\,\mathrm{MeV}$).  \label{fig:qcd_anderson_dist_D210}}
\end{figure}

\begin{figure}[t]
    \centering
     \includegraphics[width=0.49\textwidth]{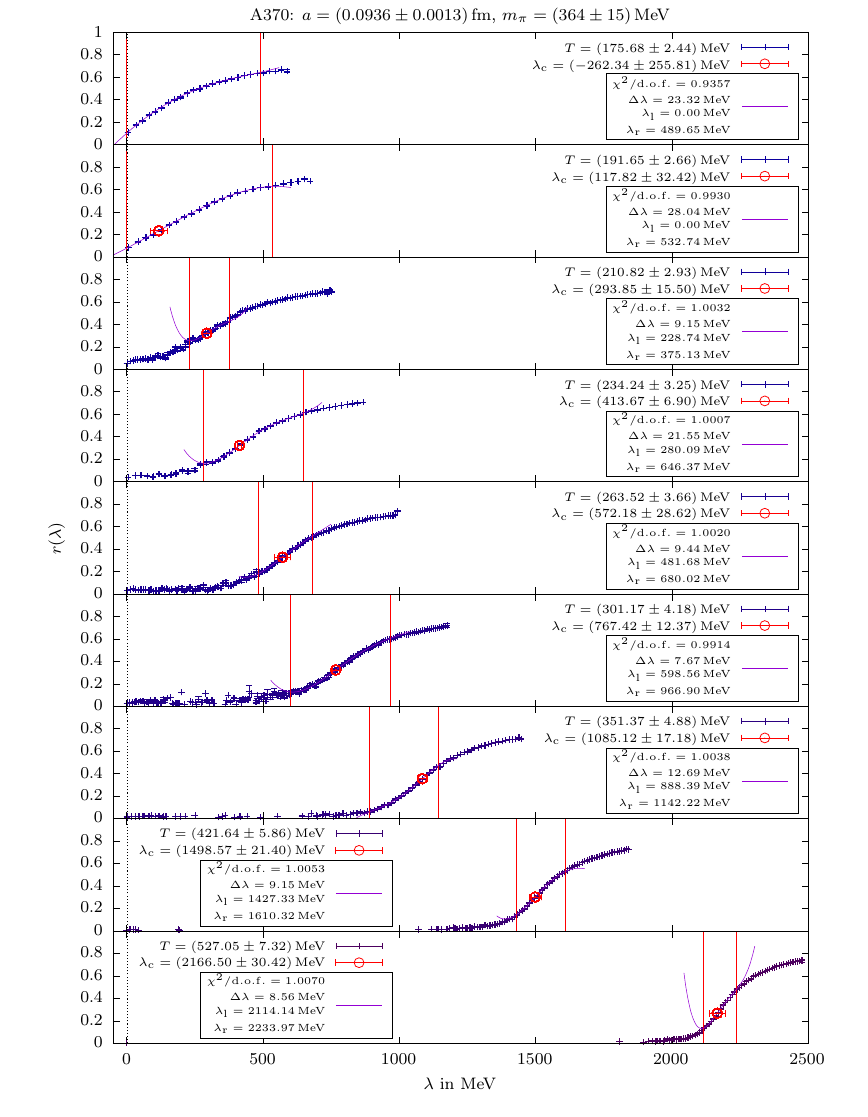}
     \caption{Bin averages of the relative eigenmode volume $r(\lambda)$ over the stereographically projected overlap eigenvalues $\lambda$, with fits~(\ref*{eq:inflectionfitl}), fit windows and extracted inflection  points marked in red, for the A370 configurations.
\label{fig:qcd_anderson_ipr_A370}}
\end{figure}

\begin{figure}[t]
    \centering
     \includegraphics[width=0.49\textwidth]{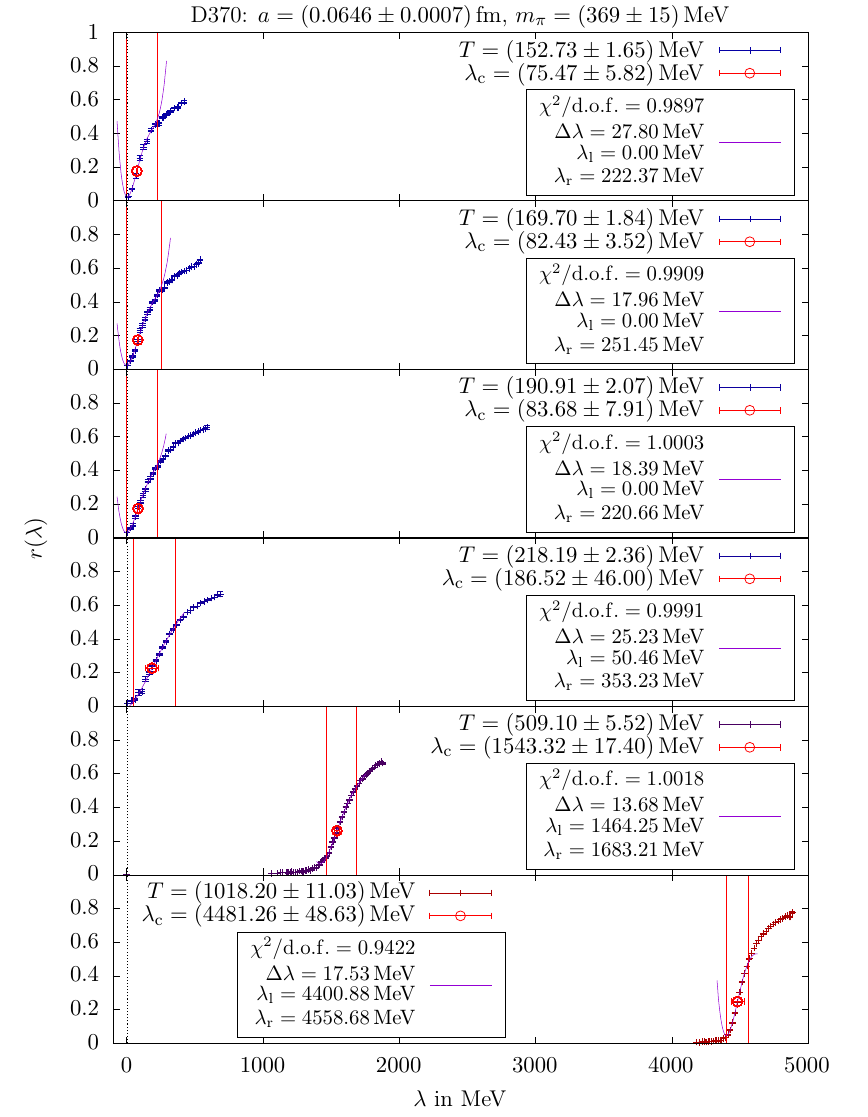}
          \caption{Same as in Fig.~\ref*{fig:qcd_anderson_ipr_A370}, here for the
     D370 configurations.
\label{fig:qcd_anderson_ipr_D370}}
\end{figure}

\begin{figure}[t]
    \centering
     \includegraphics[width=0.49\textwidth]{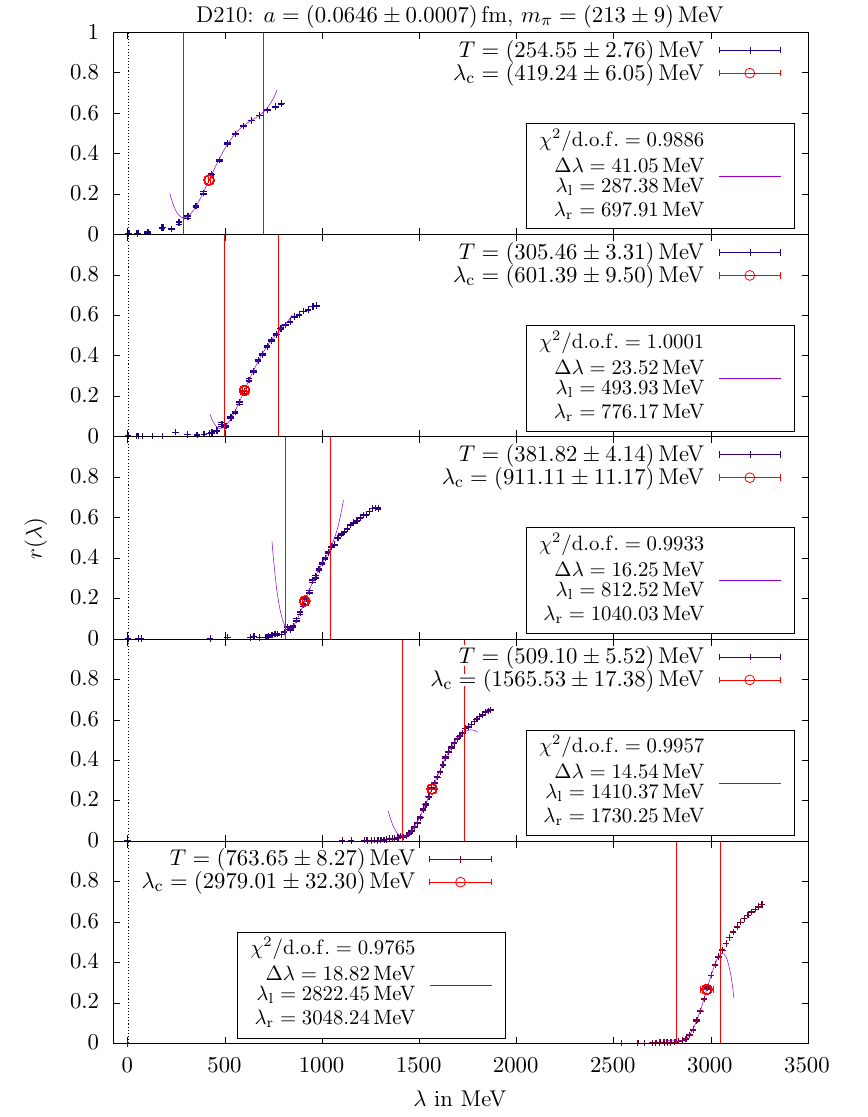}
\caption{Same as in Fig.~\ref*{fig:qcd_anderson_ipr_A370}, here for the D210 configurations.
\label{fig:qcd_anderson_ipr_D210}}
\end{figure}

\begin{figure}
\centering
\includegraphics[width=0.5\textwidth]{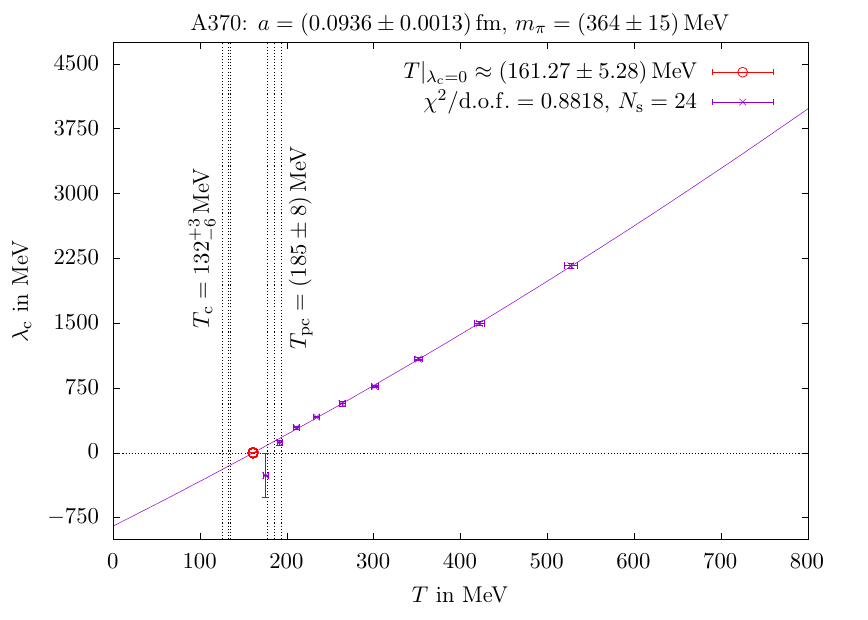}\\
\vspace{2mm}
\includegraphics[width=0.5\textwidth]
{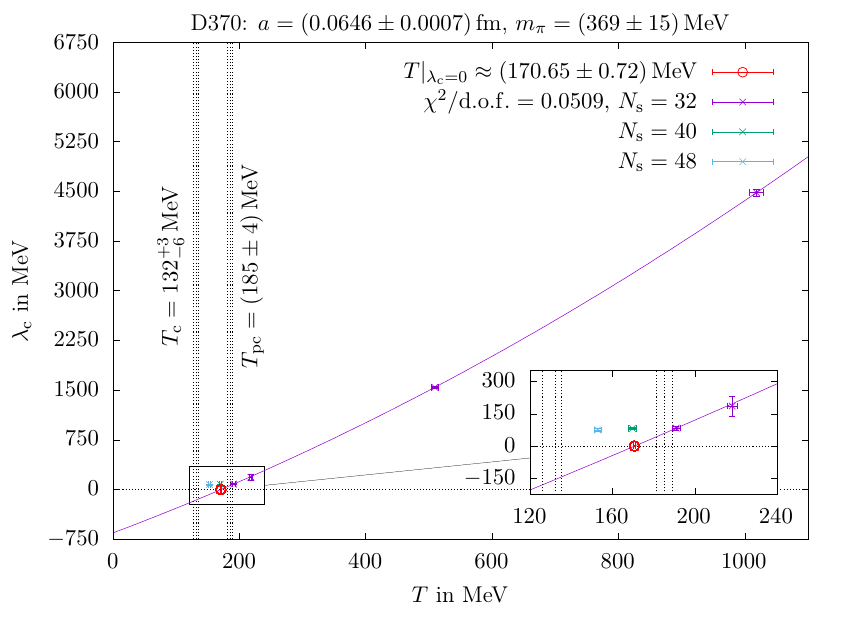}\\
\vspace{2mm}
\includegraphics[width=0.5\textwidth]{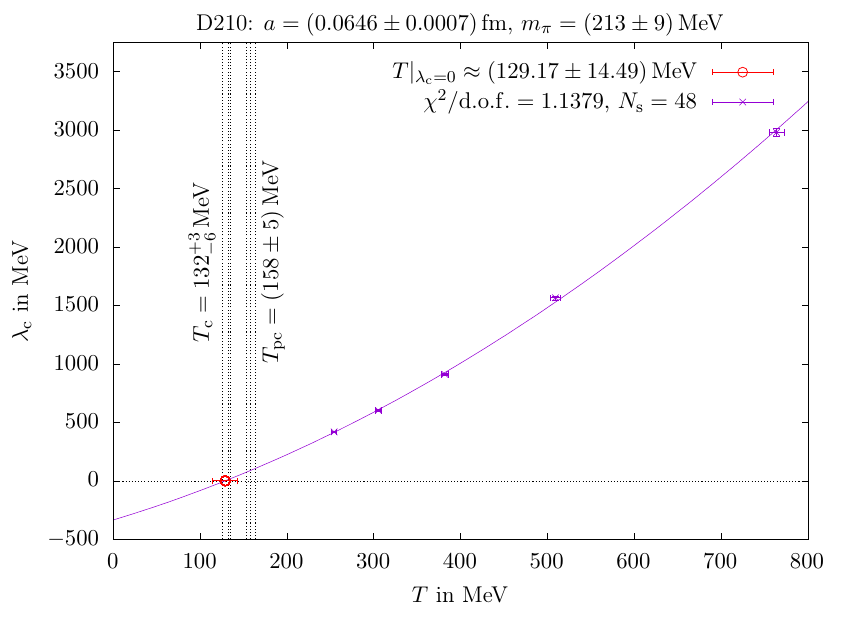}\\
\caption{Temperature dependence of mobility edge $\lambda_\mathrm{c}(T)$ for the A370 (top), D370 (middle) and D210 (bottom) configurations, as extracted from the bin-averaged relative eigenmode volumes of the stereographically projected eigenvalues in Figs.~\ref*{fig:qcd_anderson_ipr_A370},  \ref*{fig:qcd_anderson_ipr_D370} and \ref*{fig:qcd_anderson_ipr_D210}, with fits (\ref*{eq:MobilityEdgeFitp}) and $T_0$ extrapolations (red) where $\lambda_\mathrm{c}(T_0) = 0$.
} 
\label{fig:qcd_anderson_mobility_edge_over_temperature}
\end{figure}

Following the same steps as before, we start with illustrating the raw eigenvalue spectrum of $\rho(\lambda)$ for all data sets in Figs. \ref{fig:qcd_anderson_dist_A370}, 
\ref{fig:qcd_anderson_dist_D370} and
\ref{fig:qcd_anderson_dist_D210}, using histograms with bin size $\tilde a\Delta\lambda = 2.5\cdot10^{-3}$.
We show the corresponding relative eigenmode volume data in Figs.~\ref{fig:qcd_anderson_ipr_A370},
\ref{fig:qcd_anderson_ipr_D370} and
\ref{fig:qcd_anderson_ipr_D210}, with fits, fit windows and resulting inflection points $\lambda_\mathrm c$ indicated in the same way as in Figs.~\ref{fig:qcd_anderson_ipr_A370_angle},  \ref{fig:qcd_anderson_ipr_D370_angle} and \ref{fig:qcd_anderson_ipr_D210_angle} above.
Here we use fits of the form
\begin{equation}
 r(\lambda)= r_\mathrm{c}+b(\lambda-\lambda_\mathrm{c})+c(\lambda-\lambda_\mathrm{c})^3+d(\lambda-\lambda_\mathrm{c})^4
 \label{eq:inflectionfitl}
		\end{equation} 
to obtain the mobility edges $\lambda_\mathrm{c}$ and again optimize the fit boundaries and bin sizes for this purpose. Finally, we show the temperature dependence $\lambda_\mathrm{c}(T)$ in Fig.~\ref{fig:qcd_anderson_mobility_edge_over_temperature}, where we again apply fits of the form
\begin{equation}
		    \lambda_\mathrm{c}(T)=b(T-T_0)+c(T-T_0)^2\,,
      \label{eq:MobilityEdgeFitp}
		\end{equation}
to obtain the Anderson transition temperatures $T_0$.
The corresponding numerical data is shown in Table \ref{tab:qcd_anderson_lambdac_from_inflection_point}.

\begin{table}
	\centering
	\begin{tabular}{|c|c|c|c|c|c|}
	\hline
Set of ensembles &	$N_\mathrm{s}$   &   $N_\mathrm{t}$	&	$T$ / MeV	&  $T/T_\mathrm{pc}$	&  $\lambda_\mathrm{c}$ / MeV \\
	\hline
 \multirow{9}{*}{\makecell{\textbf{A370}\\$T_0 = 161(5)\,\mathrm{MeV}$}}
	& \multirow{9}{*}{24}
	& 4		&	527(7)	&  2.85(13)  &  2167(30)		\\
	&	& 5		&	422(6)	&  2.28(10)  &	1499(21)		\\
	&	& 6		&	351(5)	&  1.90(9)   &	1085(17)		    \\
	&	& 7		&	301(4)	&  1.63(7)   &	767(12)		    \\
	&	& 8		&	264(4)	&  1.42(6)   &	572(29)		    \\
	&	& 9		&	234(3)	&  1.27(6)   &	414(7)		    \\
	&	& 10	&	211(3)	&  1.14(5)   &	294(16)		    \\
	&	& 11	&	192(3)	&  1.04(5)   &	118(32)		    \\
	&	& 12	&	176(2)	&  0.95(4)   &	-262(256)		\\
	\hline
	\multirow{6}{*}{\makecell{\textbf{D370}\\$T_0 = 171(1)\,\mathrm{MeV}$}}
	&\multirow{4}{*}{32}
	  &	 3	&	1018(11) &	5.50(13)  &  4481(49)	  \\
	&	  &	 6	&	509(6)	 &	2.75(7)   &  1543(17)	    \\
	&	  &	14	&	218(2)	 &	1.18(3)   &  187(46)	    \\
	&	  &	16	&	191(2)	 &	1.03(2)   &  84(8)	        \\
	\cline{2-6}
	&  40 & 18	&	170(2)	 &	0.92(2)   &  82(4)   \\
    \cline{2-6}
	&  48 &	20	&	153(2)	 &	0.83(2)   &  75(6)	        \\
	\hline
	\multirow{5}{*}{\makecell{\textbf{D210}\\$T_0 = 129(14)\,\mathrm{MeV}$}}
	&\multirow{5}{*}{48}
	 &  4  & 764(8)  &  4.83(16)  &  2979(32)             \\
  &  &  6  & 509(6)  &  3.22(11)  &  1566(17)             \\
  &  &  8  & 382(4)  &  2.42(8)   &  911(11)               \\
  &  & 10  & 305(3)  &  1.93(6)   &  601(10)               \\
  &  & 12  & 255(3)  &  1.61(5)   &  419(6)               \\
\hline
\end{tabular}
\caption{Mobility edges $\lambda_\mathrm{c}$ determined from the inflection points in our fits~(\ref*{eq:inflectionfitl}) to the bin-averaged relative eigenmode volumes $r(\lambda)$ of the stereographically projected overlap eigenvalues in Figs.~\ref*{fig:qcd_anderson_ipr_A370},  \ref*{fig:qcd_anderson_ipr_D370} and \ref*{fig:qcd_anderson_ipr_D210}, together with the extrapolated $T_0$ values from Fig.~\ref*{fig:qcd_anderson_mobility_edge_over_temperature}.}
\label{tab:qcd_anderson_lambdac_from_inflection_point}
\end{table}

Comparing Figs.~\ref{fig:qcd_anderson_mobility_edge_over_temperature_angle} and \ref{fig:qcd_anderson_mobility_edge_over_temperature},
the main observation here is that the projection appears to have a significant effect on the mobility edge only for the D210 data set. For A370 and D370, on the other hand, we find that the extrapolated $T_0$ values agree within errors with the non-projected results of the last section, and hence also fall just slightly below the pseudocritical temperatures $T_\mathrm{pc}$ for each case. In contrast, for the D210 ensemble, with the closest to physical pion mass (but smallest $m_\pi L\approx 3.34 $), the $T_0$ prediction moves substantially, from $140(10)\,\mathrm{MeV}$ down to
$129(14)\,\mathrm{MeV}$, after applying the projection. 

This new result then falls directly into the error band of $T_\mathrm{c}=132^{+3}_{-6}\,\mathrm{MeV}$ for the chiral phase transition temperature \cite{HotQCD:2019xnw}, but places  the pseudocritical crossover temperature  $T_\mathrm{pc}=158(5)\,\mathrm{MeV}$ at a distance of almost two standard deviations, see Fig.~\ref{fig:qcd_anderson_mobility_edge_over_temperature}.

\subsection{Discussion}
\label{sec:discussion}

It seems plausible that the mobility edge vanishes right at $T_\mathrm{c}$ in the chiral limit. The interesting question is what happens in the region of nearly physical quark masses. Our extrapolations suggest that the mobility edge then also vanishes near this same temperature, 
i.e.~that $\lambda_\mathrm{c}\to 0$ for $T \to T_\mathrm{c}$ from above, or at least close to that. It is possible that this transition stays put as a geometric phase transition without thermodynamic singularity, even when a non-zero quark mass acts as an explicit chiral symmetry breaking external field to turn the chiral transition into an analytic crossover. If continuous, such a geometric phase transition should then also follow a scaling law,
\begin{equation}
    \lambda_\mathrm{c}(T) \simeq b (T-T_0)^\nu \, , \label{eq:critscale}
\end{equation}
for $T \gtrsim T_0 $. We have attempted such fits with parameters $b$, $T_0$, and $\nu$ on our results for the temperature dependent  mobility edge in the D210 ensemble, i.e.~to the data from the bottom panels in Figs.~\ref{fig:qcd_anderson_mobility_edge_over_temperature_angle} and \ref{fig:qcd_anderson_mobility_edge_over_temperature}, where there is significant deviation from the linear behavior. Depending on the assumed scaling window, in particular, whether we include the highest temperature point at $T\simeq 764$~MeV or not, and whether we use the stereographic projection or not, we obtain exponents $\nu $ between 1.26(6) and 1.56(12), which would be in rough agreement with the values reported for the localization length exponent $\nu $ in three-dimensional Anderson models,  e.g.~with $\nu\simeq 1.437  $ for the unitary Anderson model \cite{2015PhRvB..91r4206U}. The quality of the fits is generally worse than the ones presented in the previous subsections, however, and considerable spread of the $T_0$ values results in a range of temperatures below $T_\mathrm{c}$. We conclude that we can currently not reliably extract this kind of critical behavior from our data.

Alternative scenarios could follow from the existence of the IR phase with scale invariant infrared behavior in QCD, as proposed in Refs.~\cite{Alexandru:2021pap,Alexandru:2019gdm}:
Above a transition temperature in the range $200~\text{MeV} < T_\mathrm{IR} < 250~\text{MeV} $, 
the proliferation of deep infrared near-zero
modes with $0\lesssim \lambda \ll T$ in this IR phase gives rise to a second, singular mobility edge at $\lambda_\mathrm{IR}=0$. The localization properties of the deep infrared modes with $\lambda\gtrsim \lambda_\mathrm{IR}$ were found to be very similar to those with $\lambda \lesssim \lambda_\mathrm{c}$ close to the mobility edge in the bulk studied here.
To reconcile the existence of the IR phase  with the standard view on the QCD Anderson transition, it was demonstrated in \cite{Alexandru:2021xoi} that $\lambda_\mathrm{c}(T) >0  $ comes along with 
the singular mobility edge $\lambda_\mathrm{IR}=0$ at $T\gtrsim T_\mathrm{IR}$. The fate of $\lambda_\mathrm{c}(T) $ for $T\lesssim T_\mathrm{IR}$ seems less clear at the moment, however.

One speculation could be that $\lambda_\mathrm{c}(T)$ jumps down discontinuously  to $\lambda_\mathrm{IR}=0$ as $T \to  T_\mathrm{IR} $ from above. We can see that our estimates of $\lambda_\mathrm{c}(T) $ from the D210 ensemble at the moment stop right above the temperature range proposed for $T_\mathrm{IR}$.
The sudden drop of $\lambda_\mathrm{c}(T)$ near $T\sim T_\mathrm{pc}$ observed on the coarse A370 lattice with heavy pions, see the top panels in Figs~\ref{fig:qcd_anderson_mobility_edge_over_temperature_angle} and \ref{fig:qcd_anderson_mobility_edge_over_temperature}, might be seen as an indication of such a jump, but this seems somewhat far-fetched at the moment. Moreover, the volume dependence on the finer D370 lattice with comparable pion mass, shown in the inserts of the middle panels of  Figs~\ref{fig:qcd_anderson_mobility_edge_over_temperature_angle} and \ref{fig:qcd_anderson_mobility_edge_over_temperature}, appear to point in a different direction.

Neither can we currently rule out that there is an annihilation of the two mobility edges $\lambda_\mathrm{c}$ and $\lambda_\mathrm{IR}$ at any point between the extrapolated $T_0$ estimate here and the transition temperature $T_\mathrm{IR}$, below which $\lambda_\mathrm{IR}(T)$ might rise from the singular edge at zero, and intersect with our $\lambda_\mathrm{c}(T) $ curves. At such an intersection point, the band of localization in the spectrum of the QCD Dirac operator between $\lambda_\mathrm{IR}$ and $\lambda_\mathrm{c}$ would then close continuously for some temperature between $T_\mathrm{c}$ and $T_\mathrm{IR}$. Indications that this might  happen with 2+1 flavors of Highly Improved Staggered Quarks at $T \simeq 1.15\,  T_\mathrm{pc}$ in the continuum limit were recently in fact given in Ref.~\cite{Kaczmarek:2023bxb}. In either case, this would then of course invalidate extrapolations of $\lambda_\mathrm{c}(T)$ down to $\lambda_\mathrm{c}(T_0) = 0 $, in the crossover region around the physical quark-mass values.

One might nevertheless expect that in the chiral limit, the chiral phase transition coincides with the transition into the IR phase \emph{and} the merging of the mobility edges at zero, i.e.~that $T_\mathrm{c} = T_0 = T_\mathrm{IR}$, at least in this limit.  


\section{Conclusion and Outlook}\label{sec:conclusion}

In this work, we have studied the spectral properties of the QCD overlap operator in the high-temperature phase on $\mathrm{SU}(3)$ gauge-field configurations created with
dynamical twisted-mass Wilson quarks and the Iwasaki gauge action, for several temperatures, with 
two different lattice spacings, two essentially different (physical) volumes and two different pion masses. 
After presenting the raw eigenvalue spectrum and demonstrating the 
formation of the Banks-Casher gap, we have extracted the relative volume that is occupied by eigenmodes from the inverse participation ratio across a broad spectral range, and obtained estimates for the mobility edge from the inflection points. We have finally discussed the temperature dependence of the mobility edge $\lambda_\mathrm c(T)$ for all data sets. 

While we have confirmed the linear $T$-dependence seen in earlier work for the configurations in smaller volumes with coarser lattice spacings and larger pion masses, we find that a quadratic contribution appears when moving towards the continuum limit and the physical point. This has been demonstrated with high statistical significance. Furthermore, when attempting to extrapolate the mobility edge towards the chiral transition, we find that the temperature $T_0 $ where the mobility edge vanishes, $\lambda_\mathrm c(T_0)=0$, consistently undershoots the chiral crossover temperature $T_\mathrm{pc}$ predicted for the given value of the pion mass of each data set. This effect becomes more pronounced as lattice artifacts get under better control. For our best data set (with the largest physical volume, smallest lattice spacing and smallest pion mass), we find that the Anderson transition falls much closer to the predicted critical temperature $T_\mathrm c$ of the chiral phase transition, in the limit of vanishing quark masses. In fact, we observe a nearly perfect coincidence, within statistical errors, after applying a stereographic projection of the eigenvalues onto the imaginary axis, to mimic the continuum limit.

As we have discussed in Sec.~\ref{sec:discussion}, the presence of a singular second mobility edge $\lambda_\mathrm{IR}=0 $ in the IR phase of Refs.~\cite{Alexandru:2019gdm,Alexandru:2021pap} can potentially invalidate extrapolating $\lambda_\mathrm c(T) $ to zero. It will therefore be important to analyse larger $N_\mathrm t $ lattices for lower temperatures also in the D210 ensembles, which is possible in principle, although numerically increasingly expensive. This could allow distinguishing critical scaling near $T_0$ as in Eq.~\ref{eq:critscale} from the alternative scenarios  related the transition into the IR phase as discussed above. Several different spatial volumes with otherwise identical lattice parameters would furthermore allow proper finite-size analyses. Instead of the generalised participation ratios one could then, in particular, also study the effective infrared dimensions $d_\mathrm{IR} $ introduced in Ref.~\cite{Horvath:2018aap} from the volume dependence of an effective (measure-based) mode volume to classify localized, delocalized and critical modes.

With the framework created for this project, there is a wide range of other possible topics for future studies. At present, there are ongoing efforts to study the \emph{unfolded level spacing distribution} (ULSD) for the different data sets. It is known that electron-mode delocalization is reflected in the ULSD for each of the different variants of Anderson's model, which are classified by the symmetries of the Hamiltonian \cite{2015PhRvB..91r4206U}. 
The delocalized modes are thereby well-described with Wigner-Dyson random matrix theories (RMT) \cite{Evangelou1996}, while the level spacings for localized modes are Poisson distributed \cite{GarciaGarcia:2006gr,Kovacs:2010wx,Kovacs:2012zq}. 

QCD and related theories appear to follow a similar pattern: in the chirally broken phase, the entire Dirac operator spectrum follows RMT predictions \cite{Verbaarschot:2000dy,Klein:2000pj}. Different random matrix ensembles
are thereby realized for different gauge groups: the chiral Gaussian unitary 
ensemble (GUE) with Dyson-index $\beta_\mathrm{D}=2$ for three-color QCD~\cite{Kovacs:2012zq}; the Gaussian orthogonal ensemble (GOE) with $\beta_\mathrm{D}=1$ 
for two-color QCD in the continuum \cite{Wilhelm:2019fvp};
the Gaussian symplectic ensemble (GSE) with $\beta_\mathrm{D}=4$ in any-color QCD 
with quarks in the adjoint representation \cite{Hands:2000ei}, fundamental quarks in $\mathrm{G}_2$-QCD 
\cite{Wellegehausen:2013cya} or QC$_2$D with staggered quarks in the bulk phase
\cite{Halasz:1995vd,Kogut:2003ju}. 
Using the same setup as in this work, it should be possible to study how the ULSD of the localized ($\lambda < \lambda_\mathrm{c}$) modes in the chirally restored phase changes to Poissonian as it does for staggered quarks. 

Also ongoing is a study of the effect of UV-smoothing the gauge configurations with gradient flow, in order to verify, e.g., that the anti-correlation between the Polyakov loop and the scalar density of eigenmodes, as discussed in 
several previous works \cite{Bruckmann:2011cc,Cossu:2016scb,Holicki:2018sms,Baranka:2021san,Baranka:2022dib}, is preserved after smoothing. 
Local Polyakov loop fluctuations in an environment of predominantly ordered ones are believed to provide favorable regions for low Dirac modes in what has become known in the literature as the sea/islands picture which possibly connects deconfinement to the chiral symmetry restoration.

Future studies might revisit the connection between mode localization and the
topological structure of QCD \cite{Cossu:2016scb}, i.e.~by studying the  topological
charge density and its correlation to the scalar density of eigenmodes.  
Model studies have proposed that zero modes are bound to instantons and overlaps between neighboring 
instantons create a chiral condensate as the zero modes delocalize and obtain 
finite eigenvalues \cite{GarciaGarcia:2005dp,GarciaGarcia:2005vj,GarciaGarcia:2006gr}. 
While the effectiveness of this mechanism has been questioned \cite{Kovacs:2017uiz},
this is possibly a driver of the Anderson transition in QCD. 
Modes close to $\lambda_\mathrm{c}$ have also been shown to possess a multifractal structure in previous works 
\cite{Nishigaki:2013uya,Giordano:2014pfa,Ujfalusi:2015nha},
and the mobility edge
was identified as a scale-invariant fixed point of the 
ULSD in QCD with two and three colors \cite{Nishigaki:2013uya}. Both of these issues are of great interest, and deserve closer inspection.

\begin{acknowledgments}
D.~S.\ received funding from the European Union's Horizon 2020 research and innovation program under grant agreement No.~871072.
Early parts of this work were also supported by the Helmholtz International Center for FAIR within the LOEWE initiative of the State of Hesse. 
We thank Michael Ilgenfritz and Maria Paola Lombardo for many helpful discussions and for providing the tmfT configurations. We are indebted to Lukas Holicki for an implementation of the overlap Dirac operator. Helpful discussions with Andrei Alexandru, Matteo Giordano, Ivan Horv\'ath, 
Andrey Kotov, and Tam\'as Kov\'acs are also acknowledged. We are especially grateful to Ivan Horv\'ath and Matteo Giordano for communicating their helpful comments on a previous version of our manuscript.
Computational resources were provided by the HPC Core Facility and the clusters of the Institute for Theoretical Physics with support of the HRZ of Justus-Liebig University Giessen.
\end{acknowledgments}

\begin{figure*}
   \centering                              
    \includegraphics[width=0.47\textwidth]{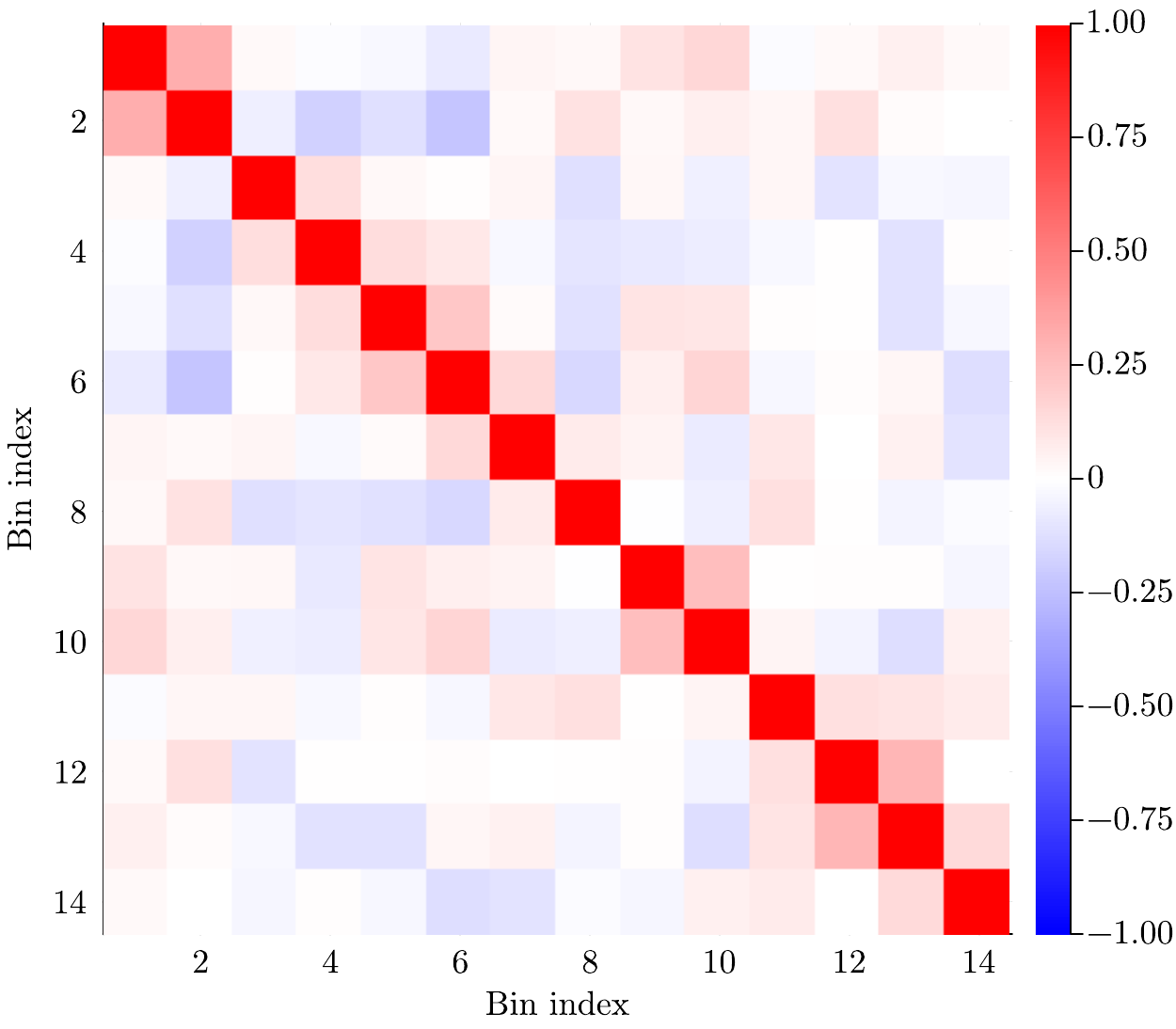} \hfill
    \includegraphics[width=0.47\textwidth]{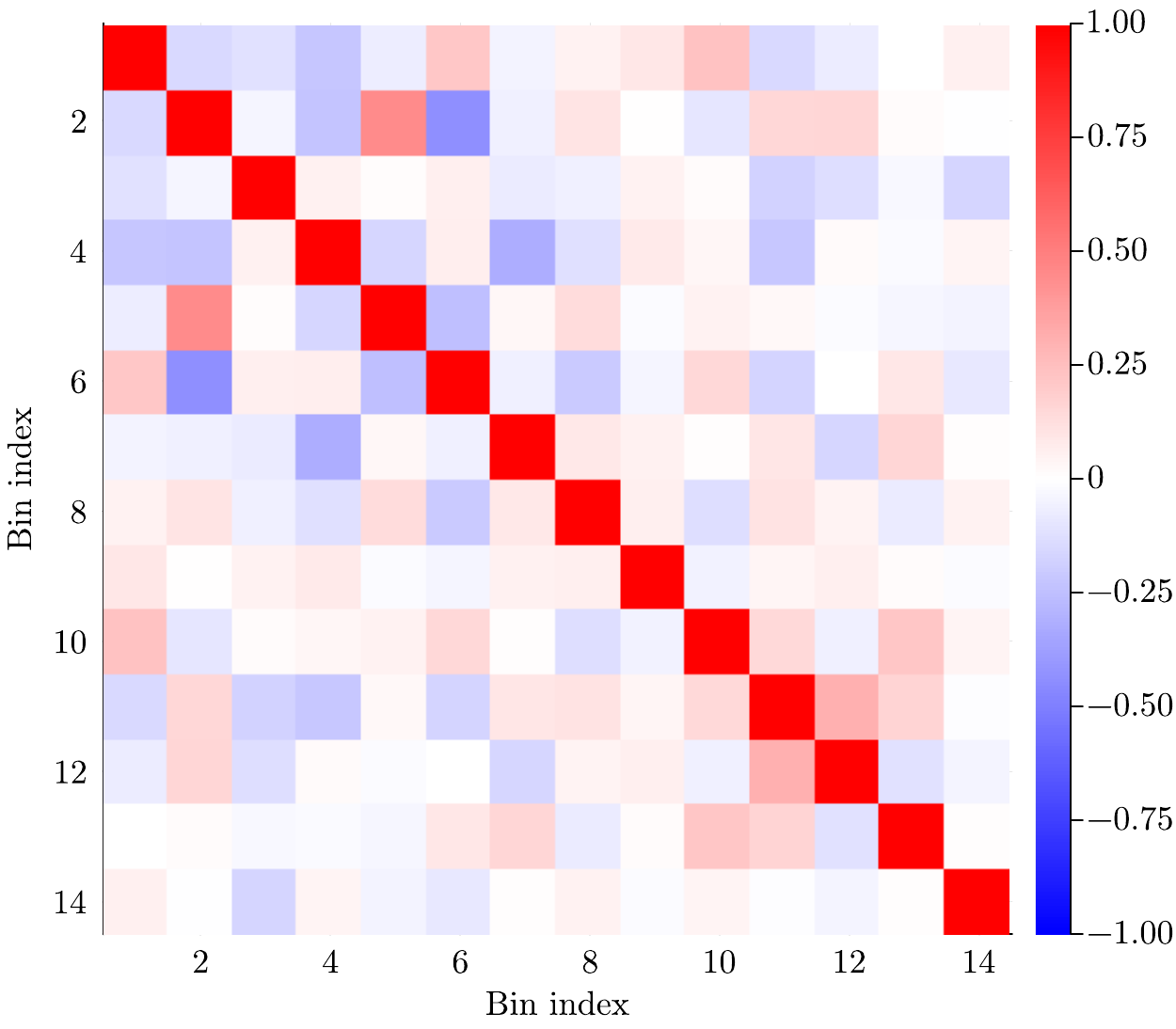}\\[8pt]
    \includegraphics[width=0.47\textwidth]{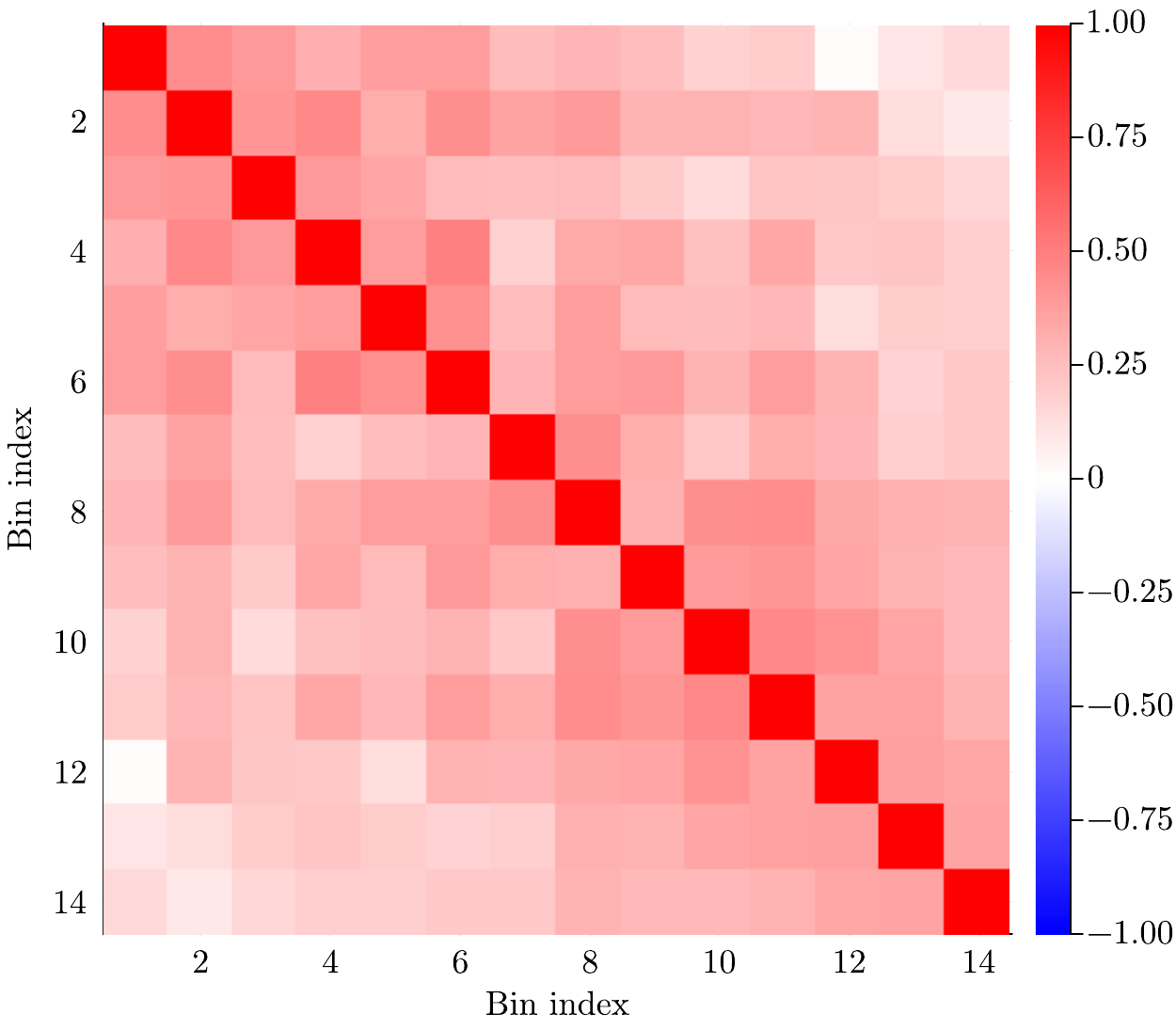}\hfill
    \includegraphics[width=0.47\textwidth]{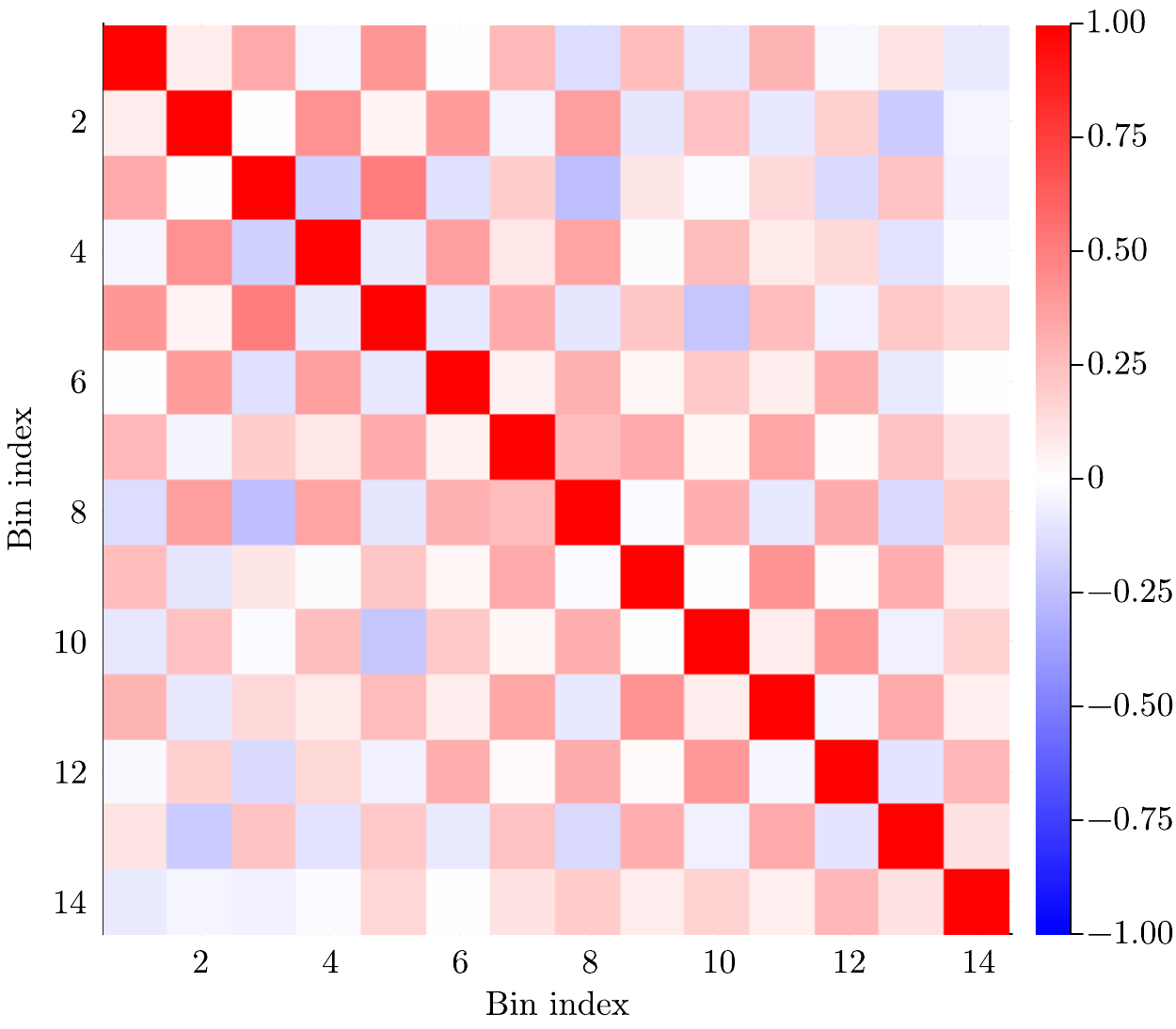}
 	\caption{Left: Correlation matrices of the stereographically projected eigenvalues (top) and relative volumes (bottom) between all bins within the fit window for $N_\mathrm{t}=4$ of the A370 data set. Right: Same correlation matrices with correlations between neighboring bins removed when using the alternating configurations  method. } 
	\label{fig:correlation}
\end{figure*}

\begin{table}
	\centering
	\begin{tabular}{|c|c|c|c|c|}
	\hline
Set of ensembles &	$N_\mathrm{s}$   &   $N_\mathrm{t}$	&	$\lambda_\mathrm{c}$ / MeV   &  $\lambda_{\mathrm{c},\mathrm{a}}$ / MeV \\
	\hline
 \multirow{9}{*}{\makecell{\textbf{A370}\\$T_0 = 161(5)\,\mathrm{MeV}$\\$T_{0,\mathrm{a}} = 158(6)\,\mathrm{MeV}$}}
	& \multirow{9}{*}{24}
	& 4		&	2167(30)	&  2182(31)		\\
	&	& 5		&	1499(21)	&	1488(21)		\\
	&	& 6		&	1085(17)	&	1055(15)		\\
	&	& 7		&	767(12)	&	759(15)		    \\
	&	& 8		&	572(29)	&	578(9)		    \\
	&	& 9		&	414(7)	&	404(7)		    \\
	&	& 10	&	294(16)	&	303(7)		    \\
	&	& 11	&	118(32)	&	122(25)		    \\
	&	& 12	&	-262(256)	&	33(67)		\\
	\hline
	\multirow{6}{*}{\makecell{\textbf{D370}\\$T_0 = 171(1)\,\mathrm{MeV}$\\$T_{0,\mathrm{a}} = 168(2)\,\mathrm{MeV}$}}
	&\multirow{4}{*}{32}
	  &	 3	&	4481(49) &  4480(49)	  \\
	&	  &	 6	&	1543(17)	 &  1546(17)	    \\
	&	  &	14	&	187(46)	 &  194(20)	    \\
	&	  &	16	&	84(8)	 &  97(6)	        \\
	\cline{2-5}
	&  40 & 18	&	82(4)	 &  91(12)   \\
    \cline{2-5}
	&  48 &	20	&	75(6)	 &  82(16)	        \\
	\hline
	\multirow{5}{*}{\makecell{\textbf{D210}\\$T_0 = 129(14)\,\mathrm{MeV}$\\$T_{0,\mathrm{a}} = 137(9)\,\mathrm{MeV}$}}
	&\multirow{5}{*}{48}
	 &  4  & 2979(32)  &  2962(32)             \\
  &  &  6  & 1566(17)  &  1568(19)             \\
  &  &  8  & 911(11)  &  932(10)               \\
  &  & 10  & 601(10)  &  617(10)               \\
  &  & 12  & 419(6)  &  421(12)               \\
\hline
\end{tabular}
\caption{Comparison of the results for the mobility edges and their extrapolations listed in Table \ref{tab:qcd_anderson_lambdac_from_inflection_point} with those for $\lambda_{\mathrm{c},\mathrm{a}}$ and $T_{0,\mathrm{a}}$ from the alternating configurations method.}
\label{tab:qcd_anderson_lambdac_from_inflection_point_alternating}
\end{table}

\appendix
\section{Correlation analysis}
\label{sec:appendix1}

Employing the criterion $\chi^2 / \mathrm{d.o.f.} \approx 1$ for fitting generally requires that the errors of all data points be statistically independent. Since, for each ensemble, we use the same configurations for every bin, the data of two different bins could in principle be correlated. In order to quantify these correlations, we have computed the matrix of correlation coefficients for each pair of bins in our fit windows. More precisely, we have computed the Pearson correlation coefficients $\rho_{X,Y} = \mathrm{cov}(X,Y) / (\sigma_X \sigma_Y)$, where the random vectors $X=(\bar{x}_1,\bar{x}_2,\bar{x}_3,\dots)$ and $Y=(\bar{y}_1,\bar{y}_2,\bar{y}_3,\dots)$ contain the configuration-wise mean values of a given observable for the respective bins. 

The left panels of Figure~\ref{fig:correlation} show the correlation matrices of the stereographically projected eigenvalue $\lambda$ as well as the relative volume $r(\lambda)$ for $N_\mathrm{t}=4$ of the A370 data set as one representative example. It indicates that the correlations of the eigenvalues are negligible, hence the respective errors can be treated as statistically independent. This is not the case for the relative volume, however, where especially neighboring bins are moderately correlated, which can be read off from the subdiagonal respectively superdiagonal, whose entries average $\sim0.39$ (same for the absolute value of the entries). 

Ideally, to remove these correlations as much as possible, one might use different configurations for every bin. This is prohibitively expensive, however.  
At least, these correlations should be drastically reduced, by sampling different configurations for neighboring bins. To test the effect of this, we have re-computed  $\rho_{X,Y}$ using data from only half of the processed configurations for bins with even index, and using data from the other half for the adjacent bins with odd index (i.e.~we alternately picked from one of the two sets
of configurations). As a result, the correlations are reduced considerably, which is demonstrated in the right panels of Figure~\ref{fig:correlation}. In this case, the subdiagonal respectively superdiagonal mean bin-bin correlation of the relative volume amounts to just $\sim0.01$ (or $\sim0.08$ for the absolute value of the entries) showing that correlations between neighboring bins have indeed been essentially removed. 

The relevant question for us is whether the inflection points change significantly for the data with the bin-bin correlations reduces in this way. Therefore, we have repeated the analysis described in Section \ref{sec:results_angle_dependence} using the method of sampling different configurations in alternating bins. The corresponding results are listed in Table~\ref{tab:qcd_anderson_lambdac_from_inflection_point_alternating}. The results for our estimates of the mobility edges with the so reduced correlations are all consistent within errors with those presented in the main text (the error bars of all inflection points overlap for both variants and are of similar overall magnitude). This indicates that while there are correlations across bins, these appear to have no significant effect on our main results. The largest shift of $T_0$ is observed for the D210 data set, where the alternating variant introduces a large statistical error due the low number of configurations, however. We therefore conclude that it was reasonable to sample all configurations for each bin in our exploratory present study. For more precision in future, however, as processing larger numbers of configurations becomes feasible, the alternating configuration variant might become a viable alternative, beyond assessing the effect of the systematic error due to these residual correlations as done here.                


\end{document}